\newif\ifplainstyle
\plainstyletrue
\plainstylefalse

\newif\ifjhepstyle
\jhepstyletrue

\newif\ifprstyle
\prstyletrue
\prstylefalse

\ifprstyle
	\documentclass[twocolumn,nofootinbib]{revtex4-1}
\else
	\documentclass[11pt,a4paper]{article}
\fi

\ifjhepstyle
	\usepackage{jheppub}
	\usepackage{amsfonts}
	\usepackage{verbatim}
	\usepackage{float}
	\usepackage{color}
	\usepackage{array}
	\newcolumntype{C}[1]{>{\centering\arraybackslash$}p{#1}<{$}}
	\makeatletter
	\def\@fpheader{\phantom{Prepared for submission to JHEP}}
	\makeatother
\else	
 	\ifprstyle
		\usepackage{verbatim}
		\usepackage{amsmath,amsfonts,amssymb}
		\usepackage[colorlinks=true
                	,urlcolor=blue
                	,anchorcolor=blue
                	,citecolor=blue
                	,filecolor=blue
                	,linkcolor=blue
                	,menucolor=blue
                	]{hyperref}
		
	\else
            	\usepackage{verbatim}
            	\usepackage{cite}
            	\usepackage{setspace}
            	\usepackage[top=2.5cm, bottom=2.75cm, left=2.5cm, right=2.5cm]{geometry}
            	\usepackage{amsmath,amsfonts,amssymb}
            	\usepackage[colorlinks=true
            	,urlcolor=blue
            	,anchorcolor=blue
            	,citecolor=blue
            	,filecolor=blue
            	,linkcolor=blue
            	,menucolor=blue
            	,linktoc=page
            	]{hyperref}
            	\usepackage{float}
            	\restylefloat{table}
            	
            	\numberwithin{equation}{section}
	\fi
\fi

\allowdisplaybreaks

\usepackage{mathtools}
\usepackage{subcaption}
\usepackage{bbold}
\usepackage{transparent}
\usepackage[normalem]{ulem}
\usepackage[textsize=footnotesize,textwidth=1.5cm]{todonotes}

\makeatletter
\let\save@mathaccent\mathaccent
\newcommand*\if@single[3]{%
  \setbox0\hbox{${\mathaccent"0362{#1}}^H$}%
  \setbox2\hbox{${\mathaccent"0362{\kern0pt#1}}^H$}%
  \ifdim\ht0=\ht2 #3\else #2\fi
  }
\newcommand*\rel@kern[1]{\kern#1\dimexpr\macc@kerna}
\newcommand*\widebar[1]{\@ifnextchar^{{\wide@bar{#1}{0}}}{\wide@bar{#1}{1}}}
\newcommand*\wide@bar[2]{\if@single{#1}{\wide@bar@{#1}{#2}{1}}{\wide@bar@{#1}{#2}{2}}}
\newcommand*\wide@bar@[3]{%
  \begingroup
  \def\mathaccent##1##2{%
    \let\mathaccent\save@mathaccent
    \if#32 \let\macc@nucleus\first@char \fi
    \setbox\z@\hbox{$\macc@style{\macc@nucleus}_{}$}%
    \setbox\tw@\hbox{$\macc@style{\macc@nucleus}{}_{}$}%
    \dimen@\wd\tw@
    \advance\dimen@-\wd\z@
    \divide\dimen@ 3
    \@tempdima\wd\tw@
    \advance\@tempdima-\scriptspace
    \divide\@tempdima 10
    \advance\dimen@-\@tempdima
    \ifdim\dimen@>\z@ \dimen@0pt\fi
    \rel@kern{0.6}\kern-\dimen@
    \if#31
      \overline{\rel@kern{-0.6}\kern\dimen@\macc@nucleus\rel@kern{0.4}\kern\dimen@}%
      \advance\dimen@0.4\dimexpr\macc@kerna
      \let\final@kern#2%
      \ifdim\dimen@<\z@ \let\final@kern1\fi
      \if\final@kern1 \kern-\dimen@\fi
    \else
      \overline{\rel@kern{-0.6}\kern\dimen@#1}%
    \fi
  }%
  \macc@depth\@ne
  \let\math@bgroup\@empty \let\math@egroup\macc@set@skewchar
  \mathsurround\z@ \frozen@everymath{\mathgroup\macc@group\relax}%
  \macc@set@skewchar\relax
  \let\mathaccentV\macc@nested@a
  \if#31
    \macc@nested@a\relax111{#1}%
  \else
    \def\gobble@till@marker##1\endmarker{}%
    \futurelet\first@char\gobble@till@marker#1\endmarker
    \ifcat\noexpand\first@char A\else
      \def\first@char{}%
    \fi
    \macc@nested@a\relax111{\first@char}%
  \fi
  \endgroup
}
\makeatother

\newcommand{\ThisIsTheTitle}{Entanglement and chaos in warped conformal field theories} 
\newcommand{\ThisIsAuthorOne}{Luis Apolo,}
\newcommand{\ThisIsAuthorTwo}{Song He,}
\newcommand{\ThisIsAuthorThree}{Wei Song,}
\newcommand{\ThisIsAuthorFour}{Jianfei Xu}
\newcommand{\ThisIsAuthorFive}{and Junjie Zheng}
\newcommand{\ThisIsTheAffiliation}{Yau Mathematical Sciences Center, Tsinghua University, Beijing 100084, China}
\newcommand{\TheseAreTheKeywords}{}

\newcommand{\ThisIsTheAbstract}{
Various aspects of warped conformal field theories (WCFTs) are studied including entanglement entropy on excited states, the R\'enyi entropy after a local quench, and out-of-time-order four-point functions. Assuming a large central charge and dominance of the vacuum block in the conformal block expansion, (i) we calculate the single-interval entanglement entropy on an excited state, matching previous finite temperature results by changing the ensemble; and (ii) we show that WCFTs are maximally chaotic, a result that is compatible with the existence of black holes in the holographic duals. Finally, we relax the aforementioned assumptions and study the time evolution of the R\'enyi entropy after a local quench. We find that the change in the R\'enyi entropy is topological, vanishing at early and late times, and nonvanishing in between only for charged states in spectrally-flowed WCFTs. 
}

\ifjhepstyle
\title{\ThisIsTheTitle}

\author[a]{\ThisIsAuthorOne}
\author[b,c]{\ThisIsAuthorTwo}
\author[a]{\ThisIsAuthorThree}
\author[d]{\ThisIsAuthorFour}
\author[a]{\ThisIsAuthorFive}

\affiliation[a]{\ThisIsTheAffiliation}
\affiliation[b]{Max Planck Institute for Gravitational Physics, Albert Einstein Institute, M\"uhlenberg 1, 14476 Golm, Germany}
\affiliation[c]{Center for Theoretical Physics and College of Physics, Jilin University, Changchun 130012, China}
\affiliation[d]{Shing-Tung Yau Center and School of Mathematics, Southeast University, Nanjing 211189, China}

\abstract{\ThisIsTheAbstract} 

\keywords{\TheseAreTheKeywords}
\fi

\begin{document}


\ifjhepstyle
\maketitle
\flushbottom
\fi

\long\def\symfootnote[#1]#2{\begingroup%
\def\thefootnote{\fnsymbol{footnote}}\footnote[#1]{#2}\endgroup} 

\def\rednote#1{{\color{red} #1}}
\def\bluenote#1{{\color{blue} #1}}

\def\({\left (}
\def\){\right )}
\def\lb{\left [}
\def\rb{\right ]}
\def\lB{\left \{}
\def\rB{\right \}}

\def\Int#1#2{\int \textrm{d}^{#1} x \sqrt{|#2|}}
\def\Bra#1{\left\langle#1\right|} 
\def\Ket#1{\left|#1\right\rangle}
\def\BraKet#1#2{\left\langle#1|#2\right\rangle} 
\def\Vev#1{\left\langle#1\right\rangle}
\def\Vevm#1{\left\langle \Phi |#1| \Phi \right\rangle}\def\bbox{\bar{\Box}}
\def\til#1{\tilde{#1}}
\def\wtil#1{\widetilde{#1}}
\def\ph#1{\phantom{#1}}

\def\ra{\rightarrow}
\def\la{\leftarrow}
\def\lra{\leftrightarrow}
\def\p{\partial}
\def\diff{\mathrm{d}}

\def\sinh{\mathrm{sinh}}
\def\cosh{\mathrm{cosh}}
\def\tanh{\mathrm{tanh}}
\def\coth{\mathrm{coth}}
\def\sech{\mathrm{sech}}
\def\csch{\mathrm{csch}}

\def\a{\alpha}
\def\b{\beta}
\def\g{\gamma}
\def\d{\delta}
\def\e{\epsilon}
\def\ve{\varepsilon}
\def\k{\kappa}
\def\l{\lambda}
\def\n{\nabla}
\def\om{\omega}
\def\s{\sigma}
\def\t{\theta}
\def\z{\zeta}
\def\vp{\varphi}

\def\ss{\Sigma}
\def\dd{\Delta}
\def\GG{\Gamma}
\def\LL{\Lambda}
\def\tt{\Theta}

\def\A{{\cal A}}
\def\B{{\cal B}}
\def\C{{\cal C}}
\def\cE{{\cal E}}
\def\D{{\cal D}}
\def\F{{\cal F}}
\def\H{{\cal H}}
\def\I{{\cal I}}
\def\J{{\cal J}}
\def\K{{\cal K}}
\def\L{{\cal L}}
\def\N{{\cal N}}
\def\O{{\cal O}}
\def\P{{\cal P}}
\def\R{{\cal R}}
\def\cS{{\cal S}}
\def\V{{\cal V}}
\def\W{{\cal W}}
\def\X{{\cal X}}
\def\Z{{\cal Z}}

\def\mfa{\mathfrak{a}}
\def\mfb{\mathfrak{b}}
\def\mfc{\mathfrak{c}}
\def\mfd{\mathfrak{d}}

\def\we{\wedge}
\def\re{\textrm{Re}}

\def\tilw{\tilde{w}}
\def\tile{\tilde{e}}

\def\tilL{\tilde{L}}
\def\tilJ{\tilde{J}}

\def\zz{\bar z}
\def\xx{\bar x}
\def\xp{x^{+}}
\def\xm{x^{-}}

\def\bp{\bar{\p}}

\def\VirU1{Vir \times U(1)}
\def\VirSL2R{\mathrm{Vir}\otimes\widehat{\mathrm{SL}}(2,\mathbb{R})}
\def\U1{U(1)}
\def\u1{U(1)}
\def\SL2R{\widehat{\mathrm{SL}}(2,\mathbb{R})}
\def\sl2r{\mathrm{SL}(2,\mathbb{R})}
\def\by{\mathrm{BY}}

\def\RR{\mathbb{R}}

\def\tr{\mathrm{Tr}}
\def\bnabla{\overline{\nabla}}

\def\sint{\int_{\ss}}
\def\dsint{\int_{\p\ss}}
\def\hint{\int_{H}}

\newcommand{\eq}[1]{\begin{align}#1\end{align}}
\newcommand{\eqst}[1]{\begin{align*}#1\end{align*}}
\newcommand{\eqsp}[1]{\begin{equation}\begin{split}#1\end{split}\end{equation}}

\newcommand{\absq}[1]{{\textstyle\sqrt{\left |#1\right |}}}



\def\none{}


\def\notebf#1{{\bf #1}}
\def\notebfr#1{{\bf \color{red} #1}}
\def\note#1{{\color{red} #1}}

\newcommand{\wei}[2]{\textcolor{blue}{#1}\todo[color=cyan]{\scriptsize{#2}}}
\newcommand{\jianfei}[2]{\textcolor{blue}{#1}\todo[color=orange]{\scriptsize{#2}}}
\newcommand{\junjie}[2]{\textcolor{blue}{#1}\todo[color=green]{\scriptsize{#2}}}
\newcommand{\song}[2]{\textcolor{blue}{#1}\todo[color=yellow]{\scriptsize{#2}}}
\newcommand{\lui}[2]{\textcolor{blue}{#1}\todo[color=red]{\scriptsize{#2}}}

\ifprstyle
\title{\ThisIsTheTitle}

\author{\ThisIsAuthorOne}
\email{\ThisIsEmailOne}

\author{\ThisIsAuthorTwo}
\email{\ThisIsEmailTwo}

\affiliation{\ThisIsTheAffiliation}


\begin{abstract}
\ThisIsTheAbstract
\end{abstract}


\maketitle

\fi

\ifplainstyle
\begin{titlepage}
\begin{center}

\ph{.}

\vskip 4 cm

{\Large \bf \ThisIsTheTitle}

\vskip 1 cm

\renewcommand*{\thefootnote}{\fnsymbol{footnote}}

{{\ThisIsAuthorOne}\footnote{\ThisIsEmailOne} } and {\ThisIsAuthorTwo}\footnote{\ThisIsEmailTwo}}

\renewcommand*{\thefootnote}{\arabic{footnote}}

\setcounter{footnote}{0}

\vskip .75 cm

{\em \ThisIsTheAffiliation}

\end{center}

\vskip 1.25 cm

\begin{abstract}
\noindent \ThisIsTheAbstract
\end{abstract}

\end{titlepage}

\newpage

\fi

\ifplainstyle
\tableofcontents
\noindent\hrulefill
\bigskip
\fi

\section{Introduction and summary} \label{se:intro}

Warped conformal field theories (WCFTs) are two-dimensional quantum field theories invariant under warped conformal transformations, spacetime symmetries where the left and right-moving coordinates $x^+$ and $x^-$  transform as~\cite{Hofman:2011zj,Detournay:2012pc}
  \eq{
  {x}^{+}{}' = f(x^+), \qquad {x}^{-}{}' = x^{-} + g(x^+), \label{transformation}
  }
and $f(x^+)$ and $g(x^+)$ are two arbitrary functions. Unlike standard conformal transformations, the warped conformal symmetries are generated by a left-moving Virasoro-Kac-Moody algebra. Notably, the $SL(2,R)_L \times U(1)_R$ global symmetries of the theory feature translations and chiral scale transformations, but do not include Lorentz transformations. Known examples of WCFT include chiral Liouville gravity~\cite{Compere:2013aya}, free Weyl fermions~\cite{Hofman:2014loa}, and free scalars~\cite{Jensen:2017tnb}. Furthermore, the Sachdev-Ye-Kitaev model with complex fermions~\cite{Davison:2016ngz} has been argued to be a broken-symmetry phase of WCFT~\cite{Chaturvedi:2018uov}.

The infinite-dimensional algebra and the lack of Lorentz symmetry make WCFTs useful in the study of holography beyond AdS spacetimes. In particular, both warped AdS$_3$ and the near-horizon throat of extremal Kerr black holes feature an $SL(2,R)\times U(1)$ isometry group~\cite{Anninos:2008fx,Compere:2008cv,Bardeen:1999px} that matches the global symmetries of WCFTs. By choosing appropriate boundary conditions, the asymptotic symmetries of a large class of gravitational theories are generated by a Virasoro-Kac-Moody algebra. This leads to a variety of holographic duals to WCFTs that include topologically massive gravity~\cite{Compere:2009zj}, stringy truncations that include matter~\cite{Detournay:2012dz}, Einstein gravity with Dirichlet-Neumann boundary conditions~\cite{Compere:2013bya}, and lower-spin gravity~\cite{Hofman:2014loa}. WCFTs also share common features with other Lorentz-violating theories, such as Galilean conformal field theories relevant to condensed matter physics and BMS field theories relevant to flat space holography~\cite{Barnich:2010eb,Bagchi:2010eg,Bagchi:2012cy,Bagchi:2014iea}. 

Several aspects of WCFTs have been studied in the literature that uncover a rich structure reminiscent of their CFT cousins. For example, the modular properties of WCFT partition functions lead to a Cardy-like formula that matches the Bekenstein-Hawking entropy of black holes in the corresponding holographic duals~\cite{Detournay:2012pc, Compere:2013bya,Azeyanagi:2018har}. Via the modular bootstrap~\cite{Apolo:2018eky}, it is furthermore known that WCFTs with a negative $U(1)$ level require states with purely imaginary charge, in agreement with bulk studies~\cite{Detournay:2012dz,Compere:2013bya}. Relatedly, correlation functions and the consequences of crossing symmetry are known and readily applicable to the bootstrap program~\cite{Song:2017czq}. 

One of the main features of WCFT studied in this paper is their entanglement entropy. The latter was computed for a single-interval on the vacuum state via the Rindler method in~\cite{Castro:2015csg,Song:2016gtd}. Interestingly, the entanglement entropy of WCFTs reveals hints of nonlocality along the $U(1)$ direction, a feature that is also observed in their correlation functions~\cite{Song:2017czq}. The holographic entanglement entropy has also been derived in both locally warped and unwarped AdS$_3$ spacetimes~\cite{Song:2016gtd} as well as lower spin gravity~\cite{Azeyanagi:2018har}. A novel feature of the holographic entanglement entropy is that it is determined by the minimal length between two null lines, the latter of which lie tangent to the modular flow emanating from the boundary end points~\cite{Song:2016gtd,Jiang:2017ecm}. This differs from the Ryu-Takayanagi prescription~\cite{Ryu:2006bv,Ryu:2006ef,Hubeny:2007xt} and suggests a deep connection between geometry and quantum entanglement beyond the AdS/CFT correspondence. 

In this paper, we explore further aspects of entanglement in WCFTs, including the entanglement entropy on excited states and the time evolution of the R\'enyi entropy after a local quench. Using similar techniques and out-of-time-order (OTO) correlators we also probe the chaotic behavior of large-$c$ WCFTs. It would be interesting to find a holographic description or our results, a task that is left for future work. 

Our results can be summarized as follows: (i) assuming a large central charge and dominance of the vacuum block in the conformal block expansion, we show that the entanglement entropy on a heavy state $\Ket{\psi}$ on the cylinder is given by
  \eq{
  S^{\psi}_{\A} &= -i P_0^{vac} \bar{\ell} + \mu q_{\psi} \ell -2 \big ( i \mu P_0^{vac} + 2 L_0^{vac} \big ) \log \bigg ( \frac{\til{\b}_{\psi}}{\pi\e} \,\sinh \frac{\pi \ell}{\til{\b}_{\psi}} \bigg ), \label{psieeintro}
  }
where $\ell$ and $\bar{\ell}$ denote the lengths of the subsystem along the Virasoro and $U(1)$ directions, $\mu$ is the spectral flow parameter defined in eq.~\eqref{spectralflow}, and $L_0^{vac}$ and $P_0^{vac}$ are the vacuum expectation values of $L_0$ and $P_0$ --- the zero modes of the Virasoro-Kac-Moody algebra. In particular, we note that eq.~\eqref{psieeintro} takes the same form as the entanglement entropy of WCFTs at inverse temperatures $\b$ and $\bar{\b}$ by means of the thermodynamical relations
  \eq{
  -i P_0^{vac}  \bigg ( \frac{\bar{\b}}{\b} - \frac{2\pi\mu}{\b} \bigg ) = \mu q_{\psi}, \qquad  \b = \til{\b}_{\psi} \equiv \frac{2\pi}{\sqrt{24 \til{h}_{\psi}/c -1}}, \label{microintro}
  }
where $\til{h}_{\psi} = h_{\psi} - q_{\psi}^2/k$ with $h_{\psi}$ and $q_{\psi}$ denoting the conformal weight and charge of $\Ket{\psi}$. We note that eq.~\eqref{microintro} is the transformation between the canonical and microcanonical ensembles in WCFT. Despite the additional ingredients and subtleties involved in this analysis, the result is similar to that found in CFT where the entanglement entropy on an excited state was also related to the entanglement entropy at finite temperature~\cite{Asplund:2014coa}.

(ii) For a generic WCFT, we find that the change of the R\'enyi entropy $\dd S^{(n)}_{\A}$ after the insertion of a local operator $\psi$ vanishes at early and late times, as in two-dimensional CFTs~\cite{Nozaki:2014hna,Nozaki:2014uaa,He:2014mwa,Guo:2015uwa, Chen:2015usa, Guo:2018lqq}. On the other hand, at intermediate times of order the length of the subsystem, we show that $\dd S^{(n)}_{\A}$ is generically nonzero and universal, depending on the spectral flow parameter $\mu$ of the WCFT and the charge $q_{\psi}$ of the operator $\psi$,
  \eq{
  \dd S^{(n)}_{\A} = \left \{ \begin{array}{lcl} 0, && \quad\textrm{early and late times}, \\
  2\pi \mu q_{\psi}, &&\quad \textrm{intermediate times}.
  \end{array} \right. \label{deltas2intro}
  }
A precise definition of early, late, and intermediate times will be given in eq.~\eqref{deltas2final}. This is the so-called memory effect of the R\'enyi entropy~\cite{Asplund:2015eha} which, while nonvanishing, differs from the one found in rational CFTs~\cite{Nozaki:2014hna, Nozaki:2014uaa, He:2014mwa, Guo:2015uwa, Chen:2015usa, Guo:2018lqq} and irrational CFTs~\cite{Asplund:2015eha,He:2017lrg, Kusuki:2017upd, Kusuki:2018wpa}. Relatedly, we note that $\dd S^{(n)}_{\A}$ vanishes at all times in a chiral CFT which features only one copy of the Virasoro algebra.
 
(iii) Assuming once again a large central charge and vacuum block dominance, we compute the OTO four-point function and show that WCFTs are maximally chaotic, like their CFT cousins~\cite{Roberts:2014ifa}. This result is compatible with the existence of black holes in the holographic duals to this class of WCFTs, and it further distinguishes warped from chiral CFTs, the latter of which feature no chaos~\cite{Perlmutter:2016pkf}. Interestingly, both $\dd S^{(n)}_{\A}$ and the OTO correlator exploit the analytic properties of correlation functions in different ways. We will see that $n$-point correlators in WCFT factorize into two parts whose form is determined by the global $SL(2,R)$ and $U(1)$ symmetries of the theory. In particular, we find that a nonvanishing $\dd S^{(n)}_{\A}$ relies on multivaluedness of the $U(1)$ part of correlation functions, while the OTO correlator exploits multivaluedness of the $SL(2,R)$ part instead.
 
The paper is organized as follows. In Section~\ref{se:wcft} we set up our conventions and discuss relevant aspects of WCFT including the symmetries, spectral flow, and correlation functions. The entanglement entropy on excited states is derived in Section~\ref{se:excited}. Therein the semiclassical approximation of the Virasoro-Kac-Moody vacuum block is also discussed. The time evolution of the R\'enyi entropy after a local quench is considered in Section~\ref{se:quench} while the OTO correlator and its consequences are studied in Section~\ref{se:chaos}. We derive the Virasoro-Kac-Moody block in Appendix~\ref{ap:block} and the warped conformal transformation between an $n$-sheeted Riemann surface and the complex plane in Appendix~\ref{ap:map}.


\section{Aspects of warped CFT} \label{se:wcft}  
In this section we review the salient features of WCFT that are needed in our analysis following refs.~\cite{Hofman:2011zj,Detournay:2012pc,Song:2017czq}. We begin by considering WCFTs on the ``reference plane'' where the vacuum state is left invariant under the global $SL(2,R) \times U(1)$ symmetries of the theory. We will pay special attention to the ``tilt'' $\mu$ of the WCFT which is obtained via spectral flow of the Virasoro-Kac-Moody algebra. Finally, we describe the structure of the two, three, and four-point functions and the conformal block expansion of the latter.


\subsection{Warped CFT on the reference plane}

Warped CFTs are characterized by two chiral currents $T(x^+)$ and $P(x^+)$ that generate the warped conformal transformations~\eqref{transformation}~\cite{Hofman:2011zj}. In this subsection we consider WCFTs on the reference plane where $\Vev{T(x^+)} =  \Vev{P(x^+)} = 0$.  On the reference plane the infinitesimal warped conformal transformations are generated by 
 \eq{
  l_n = - (x^+)^{n+1} \p_+, \qquad p_n = i (x^+)^n \p_-,    \label{generators}
  }
where $\p_{\pm} \equiv \p/\p x^{\pm}$. In particular, $l_0$ generates scale transformations of $x^+$ while $p_0$ is responsible for translations along $x^-$. The $U(1)$ symmetry is not necessarily compact and, as a result, the $U(1)$ charge is not necessarily quantized. 

The charges corresponding to the infinitesimal transformations~\eqref{generators} are given by the modes of the $T(x^+)$ and $P(x^+)$ currents,
   \eq{
  L_n &= -\frac{i}{2\pi} \int dx \, (x^+)^{n+1} T(x^+), \qquad P_n = -\frac{1}{2\pi} \int dx\, (x^+)^n P(x^+),
  }
which generate a chiral Virasoro-Kac-Moody algebra with central charge $c$ and level $k$,
 \eqsp{
 [L_n, L_m] &= (n-m) L_{n+m} + \frac{c}{12} n(n^2 - 1) \d_{n+m}, \\
 [L_n, P_m] & = -m P_{n+m}, \\
 [P_n, P_m] & = \frac{k}{2} n \d_{n+m}. \label{algebra}
 }
 In principle, the level of the Kac-Moody algebra can be rescaled to one. However, different sectors in some holographic WCFTs may be interpreted as having different signs of $k$ so we leave the $U(1)$ level intact~\cite{Apolo:2018eky}. 
 
On the reference plane the vacuum is invariant under the global $SL(2,R)\times U(1)$ symmetries of the WCFT and is assumed to satisfy
 \eq{ 
 L_{n} \Ket{0} = P_{n+1}\Ket{0} =0, \qquad n \ge -1.
 }
The $SL(2,R)\times U(1)$ invariance of the theory on the reference plane constrains the correlation functions of WCFTs in a similar way that the $SL(2,R)\times SL(2,R)$ group constrains correlation functions of two-dimensional CFTs~\cite{Song:2017czq}.


\subsection{Spectral flow and vacuum charges}

Besides $c$ and $k$, WCFTs are characterized by the tilt parameter $\mu$ which has the following equivalent physical interpretations: (i) as a spectral flow parameter, (ii) as the $L_0$ and $P_0$ charges of the vacuum, (iii) as the tilt of the cylinder~\cite{Detournay:2012pc}. First of all, the tilt $\mu$ can be introduced via a warped conformal transformation
  \eq{
  x^- \to y, \qquad y = x^- + \mu \log x^+. \label{spectralflow}
  }
This transformation is compatible with eq.~\eqref{transformation} and is equivalent to spectral flow. Indeed, under a finite warped conformal transformation, the $T(x^+)$ and $P(x^+)$ currents transform as~\cite{Detournay:2012pc}
  \eq{
  T'(x^+{}') & = \bigg ( \frac{\p x^+}{\p x^+{}'} \bigg)^2 \Big [ T(x^+) - \frac{c}{12} \{ x^+{}', x^+ \} \Big ] + \frac{\p x^+}{\p x^+{}'} \frac{\p x^-}{\p x^+{}'} P(x^+) - \frac{k}{4} \bigg ( \frac{\p x^-}{\p x^+{}'} \bigg )^2, \label{Tprime} \\
  P'(x^+{}') & = \bigg ( \frac{\p x^+}{\p x^+{}'} \bigg ) \Big [ P(x^+) + \frac{k}{2} \frac{\p x^-{}'}{\p x^+} \Big ], \label{Pprime}
  }
where $\{f,x\}$ is the Schwarzian derivative. It is not difficult to check that the new ${L}_n'$ and ${P}'_n$ modes satisfy
  \eq{
  {L}'_n & = L_n - i \mu P_n - \frac{\mu^2 k}{4} \d_n, \qquad {P}'_n = P_n - \frac{i\mu k}{2} \d_n, \label{spectralflowcurrents}
  }
which is the spectral flow transformation that leaves the Virasoro-Kac-Moody algebra intact. Furthermore, if we map the plane to the cylinder via the exponential map $x^+ \to e^{i x^+}$, then the transformation~\eqref{spectralflow} tilts the cylinder. 
  
In what follows we consider WCFTs with $\mu \ne 0$. Hence the coordinates of the WCFT on the (spectrally-flowed) plane are given by
  \eq{
 (x^+, y) = (x^+,\, x^- + \mu \log x^+). \label{spectralflowcoordinates}
 }
Although this is equivalent to spectral flow, a nonvanishing tilt $\mu$ carries physical consequences.\footnote{This is reminiscent of string theory on AdS$_3$ where spectral flow generates altogether new representations of the symmetry algebra~\cite{Maldacena:2000hw}.} Indeed, spectral flow affects both the entropy~\cite{Detournay:2012pc} and the entanglement entropy of WCFTs~\cite{Castro:2015csg,Song:2016gtd}, and plays an important role in the holographic derivation of the latter~\cite{Song:2016gtd}. Relatedly, we note that spectral flow leads to the following nonvanishing vacuum expectation values\footnote{The Sugawara-like combination of charges $L_0 - P_0^2/k \ge 0$ is invariant under spectral flow. The inequality is required by unitarity and is saturated by the vacuum.}
  \eq{
  \big\langle 
  {L}'_0\big\rangle= -\frac{\mu^2 k}{4}, \qquad \big\langle {P}'_0 \big\rangle = -\frac{i \mu k }{2}. \label{L0P0vevs}
  }
From this point of view, the spectral flow transformation is similar to the conformal transformation in CFT that maps the plane to the cylinder. The latter introduces a length scale $R$ --- the size of the circle that is analogous to $\mu^{-1}$ --- which leads to vacuum expectation values for the zero mode charges of the CFT. Note that if the spectral flow parameter $\mu$ is real, then $\big\langle {P}'_0 \big\rangle$ is imaginary. This is not a bug but a feature of WCFTs with holographic duals~\cite{Detournay:2012pc}. In this class of holographic WCFTs the dual theory is sensitive only to $\Vev{P_0'}^2$ and the metric and conserved charges are always real.


\subsection{Correlation functions and conformal blocks}

On the reference plane, the global $SL(2,R) \times U(1)$ symmetries of a WCFT completely fix the coordinate dependence of its two and three-point functions. On the other hand, as in standard CFT, the four-point function is determined only up to a function of the cross-ratio,
  \eq{
  z = \frac{x^+_{12} x^+_{34}}{x^+_{13} x^+_{24}}, \label{crossratio}
  }
where we use $a_{ij} \equiv a_i - a_j$ for any indexed variable $a_i$. In particular, the two, three, and four point functions are given by~\cite{Song:2017czq}
  \eq{
\Vev{\phi_1 \phi_2} &=  \Big (e^{i \sum_j q_j x^-_j} \d_{\,\ss_i q_i} \Big ) \frac{ \d_{h_1 - h_2}  }{(x^+_{12})^{2 h_1}}, \label{2pt} \\
\Vev{\phi_1 \phi_2 \phi_3} &=  \Big(  e^{i \sum_j q_j x^-_j}  \d_{\,\ss_i q_i} \Big ) \frac{C_{123}}{(x^+_{12})^{h_1+h_2-h_3} (x^+_{23})^{h_2 + h_3 - h_1} (x^+_{31})^{h_3 + h_1 - h_2}}, \label{3pt} \\
 \big\langle\phi_1 \phi_2 \phi_3 \phi_4\big\rangle &= \Big(  e^{i \sum_j q_j x^-_j}  \d_{\,\ss_i q_i}\Big )  \bigg ( \frac{x^+_{24}}{x^+_{14}} \bigg)^{ h_{12}} \bigg ( \frac{x^+_{14}}{x^+_{13}} \bigg)^{ h_{34}}  \frac{ G(z)  }{(x^+_{12})^{h_1+h_2}(x^+_{34})^{h_3+h_4}} ,   \label{4pt}
  }
where $C_{123}$ is the OPE coefficient and $\phi_i \equiv \phi_i(x^+_i,x^-_i)$ is a Virasoro-Kac-Moody primary field with conformal weight $h_i$ and $U(1)$ charge $q_i$ --- the eigenvalues of $L_0$ and $P_0$, respectively. Note that the $n$-point functions vanish unless the total momenta along $x^-$ is zero, i.e.~unless the total $U(1)$ charge vanishes
  \eq{
  \sum_i q_i = 0.
  }
The $x^+$-dependence of the $n$-point functions is the same as that of chiral CFTs with a left-moving Virasoro algebra. In contrast, the dependence on $x^-$ suggests WCFTs are nonlocal along this direction. This interpretation is supported from the single-interval entanglement entropy of WCFTs, the latter of which features hints of nonlocality along the $x^-$ coordinate~\cite{Castro:2015csg,Song:2016gtd}. A similar phenomenon is observed in the entanglement entropy on excited states derived in Section~\ref{se:excited}.

Under the warped conformal transformation~\eqref{transformation} a primary field $\phi(x^+,x^-)$ transforms as an $h$-form under Virasoro transformations and as a scalar under $U(1)$ transformations such that
  \eq{
  {\phi}'(x^+{}',x^-{}') = \bigg (\frac{\p {x}^+{}'}{\p x^+} \bigg )^{-h} \phi(x^+,x^-).
  }
This implies that the $n$-point functions transform as
  \eq{
 \!\! \Big\langle\prod_{i=1}^n {\phi}'_i(x^+_i{}',x^-_i{}') \Big\rangle = \prod_{i=1}^n \bigg (\frac{\p {x}^+_i{}'}{\p x^+_i} \bigg )^{-h_i}  \Big\langle\prod_{i=1}^n\phi_i(x^+_i,x^-_i) \Big\rangle ,\label{correlatorsprime}
  }
without any additional factors of ${x}^-{}'$ on the right hand side~\cite{Song:2017czq}. Since eq.~\eqref{spectralflow} affects only the $x^-$ coordinate we find that, after spectral flow, the $n$-point functions still transform as a scalar. In particular, in terms of the coordinates $(x^+, \, x^-)$ on the reference plane, the two, three, and four-point-point functions are given by eqs.~\eqref{2pt} --~\eqref{4pt}. One may choose to express the latter entirely in terms of the $(x^+, y)$ coordinates of the spectrally-flowed theory. In that case one simply replaces the $x^-_{i}$ coordinates featured in the $n$-point functions by
  \eq{
  x^-_{i} \to y_{i} - \mu \log {x^+_j}. \label{shift}
  }

As in CFTs, the four-point function of WCFTs can be expressed in terms of Virasoro-Kac-Moody blocks such that, in the $s$-channel~\cite{Song:2017czq},
  \eq{
  G(z) = \sum_{p} C_{p 12} C^{p}{}_{34} \V_{p}(z), \label{blockexpansion}
  }
where the sum runs over primary states with weight $h_p$ and charge $q_{p} = q_3 + q_4 = -q_1 - q_2$, and the block $\V_{p}(z)$ depends on $c$, $k$, $h_i$, $q_i$, $h_p$, and $q_{p}$. The contribution of the $U(1)$ descendant states to $\V_{p}(z)$ can be determined exactly using the Sugawara basis. In our conventions the Virasoro-Kac-Moody block is given by
  \eq{
  \V_{p}(z)= z^{\frac{(q_3 + q_4)^2}{k}} (1-z)^{\frac{2q_2 q_3}{k}} {\F} (c-1, \til{h}_i, \til{h}_{p},z), \label{block}
  }
where $\til{h}_i = h_i - q_i^2/k$ is the conformal weight in the Sugawara basis and ${\F}(c, h_i, h_p,z)$ is the standard Virasoro conformal block. Up to conventions, eq.~\eqref{block} is the same expression found in CFTs with an internal $U(1)$ symmetry in~\cite{Fitzpatrick:2015zha}. The derivation of eq.~\eqref{block} in WCFT is given in Appendix~\ref{ap:block}. The Virasoro conformal block $\F$ can be expanded near $z = 0$ as
  \eq{
  {\F} (c-1, \til{h}_i, \til{h}_{p},z) = z^{\til{h}_{p}} \sum_{K=0}^{\infty} \F_K (c-1, \til{h}_i, \til{h}_{p})  z^K, \label{virasoroblock}
  }
where the $\F_K (c-1, \til{h}_i, \til{h}_{p})$ coefficients are determined by the Virasoro algebra with central charge $c - 1$. Finally, note that the weights $\til{h}_i$ and $\til{h}_p$ are invariant under spectral flow. Hence the conformal block expansion is independent of the spectral flow parameter.


\section{ Entanglement entropy on excited states} \label{se:excited}

In this section we derive the entanglement entropy on excited states in WCFTs defined on the cylinder. Our derivation is valid provided the following assumptions, which are natural from a holographic point of view, hold: (i) a large central charge $c$, (ii) dominance of the vacuum block in the conformal block expansion, and (iii) excited states whose conformal weights scale with $c$. The interpretation of our results is facilitated by comparison to the entanglement entropy of WCFTs at finite temperature to which we now turn.


\subsection{Entanglement entropy in warped CFT}

Let us begin by considering a WCFT on the spectrally-flowed plane. We define the subsystem $\A$ by the line connecting the points $(x^+_1, y_1)$ and $(x^+_2, y_2)$ which we denote by
  \eq{
  \A: \quad (x^+, y) \in \big [ (x^+_1, y_1), \, (x^+_2, y_2) \big]. \label{intervalA}
  }
The reduced density matrix on $\A$ is obtained by tracing out the contribution of the complement $\A^{c}$, namely $\rho_{\A} = \tr_{\A^c} \rho$, where $\rho = \Ket{0}\Bra{0}$ and $\Ket{0}$ is the vacuum. The moments of the reduced density matrix $\rho_{\A}$ are given by the R\'enyi entanglement entropies
  \eq{
  S^{(n)}_{\A} = \frac{1}{1-n} \log \tr \big ( \rho^n_{\A} \big ), \label{renyi}
  }
whose $n \to 1$ limit yields the von Neumann entropy of subsystem $\A$, that is
  \eq{
  S_{\A} = \lim_{n \to 1} S^{(n)}_{\A} = -\tr \rho_{\A} \log \rho_{\A}. \label{vonneumann}
  }

The R\'enyi entropies~\eqref{renyi} can be computed in the path integral formalism by gluing together $n$ replicas of the system along the interval $\A$~\cite{Calabrese:2004eu,Calabrese:2009qy} (see~\cite{Rangamani:2016dms} for a review). This defines an $n$-sheeted two-dimensional manifold $\R_n$ with cuts along $\A$ such that
 \eq{
 \tr \rho_{\A}^n = \frac{Z_n}{Z_1},
 }
where $Z_n$ is the partition function evaluated on $\R_n$. We will use this approach in the next section when we study the time evolution of the R\'enyi entropy after a local quench. Alternatively, the trace of $\rho_{\A}^n$ can be evaluated by considering $n$ copies of the original WCFT defined on the plane. In this case, the replica boundary conditions are enforced by the insertion of twist fields $\s_n$ at the endpoints of $\A$ so that
  \eq{
  \tr \rho_{\A}^n = \big\langle\s_n^{\dagger}(x^+_2, y_2) \s_n(x^+_1, y_1) \big\rangle, \label{twist2pt}
  }
where the two-point function is evaluated on the cyclic orbifold WCFT$^n/\mathbb{Z}_n$ on the plane.  

The conformal weight and $U(1)$ charge of the twist field $\s_n$ were determined in~\cite{Castro:2015csg,Song:2017czq} via a generalization of the Rindler method used in CFT~\cite{Casini:2011kv}. The twist field conformal weight $h_n$ and charge $q_n$ are given by\footnote{Note that in our conventions $q_n$ has the opposite sign from the $U(1)$ charge used in~\cite{Castro:2015csg,Song:2017czq}.}
  \eq{
  h_n = n \Big ( \frac{c}{24} + \frac{L_0^{vac}}{n^2} + \frac{i \a P_0^{vac}}{2\pi n} - \frac{\a^2 k}{16\pi^2} \Big), \qquad  q_n = \Big( P_0^{vac} + \frac{i n k \a}{4\pi} \Big), \label{twistcharges}
  }
where $L_0^{vac}$ and $P_0^{vac}$ denote the vacuum expectation values of $L_0$ and $P_0$ on the cylinder. The variable $\alpha$ in eq.~\eqref{twistcharges} was first introduced as a free parameter in ref.~\cite{Song:2016gtd} in the generalized Rindler transformation of WCFT. Its value was later determined by matching the entanglement entropy of WCFT holographically in BTZ black holes. A similar observation has also been made for warped AdS black holes in lower spin gravity in~\cite{Azeyanagi:2018har}.

We now argue that $\alpha$ should be identified with the tilt parameter $\mu$ of the WCFT. Notice that when $n \to 1$ the symmetric orbifold WCFT must reduce to the original theory which implies that $h_1 = q_1= 0$ for the corresponding twist fields. This is possible if the following relation between the vacuum charges and the parameter $\alpha$ holds
  \eq{
  L_0^{vac}  = -\frac{c}{24} -\frac{\alpha^2  k}{16\pi^2}, \qquad P_0^{vac} = -\frac{i \alpha k }{4\pi}. \label{vevs}
  }
These are precisely the spectrally-flowed charges~\eqref{L0P0vevs} defined on the cylinder provided that
  \eq{
  \mu = \frac{\a}{2\pi}.\label{mualpha}
  }
Equation~\eqref{vevs}  guarantees that $h_n$ and $q_n$ are proportional to $n - 1$ as expected,
\eq{
h_n = \frac{(n-1)}{n} \bigg [ (n+1) \frac{c}{24} - (n - 1) \frac{\mu^2 k}{4} \bigg ] , \qquad q_n = (n-1) \frac{i k \mu}{2}. 
}
We also note that eq.~\eqref{vevs} saturates the unitarity bound $L_0^{vac} - (P_0^{vac})^2 /k = -c/24$, and that the following choice of $\mu$ reproduces the values of $L_0^{vac}$ and $P_0^{vac}$ found in the holographic entanglement entropy of WCFT (where $k$ is assumed to be negative)~\cite{Song:2016gtd} 
  \eq{
  \mu =  \sqrt{-\frac{c}{6}\frac{1}{k}}  \qquad \Rightarrow \qquad L_0^{vac} = 0, \qquad P_0^{vac} = \frac{i}{2} \sqrt{-\frac{c \,k}{6}}.
  }

The entanglement entropy of subsystem $\A$ is thus given by~\cite{Song:2016gtd}
  \eq{
    S_{\A} = -i P_0^{vac} \bigg [ y_{21} - \mu \log \bigg(\frac{x^+_2}{x^+_1} \bigg) \bigg ]   -2 \big ( i \mu P_0^{vac} + 2 L_0^{vac} \big ) \log \bigg( \frac{x^+_{21}}{\e} \bigg ), \label{eeplane}
  }
where we introduced a UV regulator $\e$ and $x^{+}_{21}$, $y_{21}$ denote the lengths of the interval $\A$. Eq.~\eqref{eeplane} is valid for WCFTs defined on the plane with nonvanishing tilt parameter $\mu$. When the latter vanishes, the entanglement entropy reduces to the result derived in~\cite{Castro:2015csg}. To compute the entanglement entropy at finite temperature we can exploit the symmetries of the WCFT and map the two-point function of twist fields~\eqref{twist2pt} to the thermal cylinder. This can be accomplished by the following warped conformal transformation~\cite{Song:2016gtd}
  \eq{
  x^+ = e^{\frac{2\pi}{\b} w^+}, \qquad x^- = w^- + \bigg ( \frac{\bar{\b}}{\b} - \frac{2\pi\mu}{\b} \bigg ) w^+, \label{mapbeta}
  }
where the coordinates on the thermal cylinder satisfy $(w^+, \,w^-) \sim (w^+ + i \b, \, w^- - i \bar{\b})$. Using eq.~\eqref{correlatorsprime} we obtain
  \eq{
  S^{\b,\bar{\b}}_{\A} = -i P_0^{vac} \bigg [ \bar{\ell} + \bigg (\frac{\bar{\b}}{\b} - \frac{2\pi\mu}{\b} \bigg ) \ell \bigg ] -2 \big ( i \mu P_0^{vac} + 2 L_0^{vac} \big ) \log \bigg ( \frac{\b}{\pi\e} \, \sinh \frac{\pi \ell}{\b} \bigg ), \label{thermalee}
  }
where $\ell = |w^{+}_{21}|$ and $\bar{\ell} = |w^-_{21}|$ denote the lengths of the interval $\A$ on the $(w^+,\,w^-)$ system. 
  

\subsection{Excited states and four-point functions}

Let us now consider the entanglement entropy on an excited state. The latter is obtained from the vacuum via the state/operator correspondence
  \eq{
  \Ket{\psi} = \psi(0,0) \Ket{0},
  }
whereupon the density matrix becomes $\rho  = \Ket{\psi}\Bra{\psi}$. The R\'enyi entropy can be computed in the symmetric orbifold theory using the twist fields to enforce the replica boundary conditions. In this case the trace of $\rho_{\A}^n$ is given by a four-point function~\cite{Asplund:2014coa}
  \eq{
    S_n \equiv \tr \rho_{\A}^n & =  \Bra{\Psi} \s_n^{\dagger}(x^+_2, y_2) \s_n(x^+_1, y_1) \Ket{\Psi} , \notag\\
  & = \lim_{x^+ \to \infty} (x^+)^{2 h_{\Psi}}  \big\langle \Psi^{\dagger}(x^+,0)  \s_n^{\dagger}(x^+_2, y_2) \s_n(x^+_1, y_1) \Psi(0,0) \big \rangle, \label{twist4pt}
  }
where $\Psi$ denotes the operator in the WCFT$^n/\mathbb{Z}_n$ orbifold that is obtained from the insertion of $\psi$ in each of its copies.\footnote{In particular, this implies that $h_{\Psi} = n h_{\psi}$ and $q_{\Psi} = n q_{\psi}$ where $h_{\psi}$ is the conformal weight and $q_{\psi}$ the $U(1)$ charge of $\psi$.} As reviewed in Section~\ref{se:wcft}, the four-point function~\eqref{twist4pt} can be expressed in terms of Virasoro-Kac-Moody blocks via eq.~\eqref{blockexpansion}. In the $t$-channel the conformal block expansion is given by
  \eq{
  S_n = e^{-i q_n x^{-}_{21}} z^{2 q_{\Psi} q_n} (1-z)^{-2 h_n} \sum_p C_{p\Psi\Psi} C^p{}_{\s\s} \F (c-1, \til{h}_i, \til{h}_{p}, 1-z),    \label{tchannelee}
  }
where we have set $x^+_2 = 1$ without loss of generality, $z \equiv x^+_1$ is the cross-ratio~\eqref{crossratio}, and we recall that $\F$ denotes the Virasoro conformal block. In particular, note that eq.~\eqref{tchannelee} is expected to converge for $z \sim 1$ and that the operators exchanged in the conformal block expansion are all neutral, i.e.~$q_p = 0$. 


\subsection{Entanglement in the semiclassical limit}

In this section we consider WCFTs in the semiclassical limit defined by 
  \eq{
  c \to \infty \qquad \textrm{with} \qquad \frac{\tilde{h}_i}{c}, \, \frac{\tilde{h}_p}{c} \,\,\, \textrm{fixed}. \label{semiclassical0}
  }
 There are two motivations for this definition of the semiclassical limit. One comes from bulk intuition. For instance, in Einstein gravity with Dirichlet-Neumann boundary conditions~\cite{Compere:2013bya, Apolo:2018eky}, the spectral flow invariant weight $\tilde{h}$ is identified with the sum of mass and angular momentum, i.e.~$\tilde{h} = M + J +{c/24}$. The other motivation is that, similar to the chiral sector of CFT, the Virasoro conformal block is expected to exponentiate~\cite{Belavin:1984vu,Zamolodchikov:1987abc}
  \eq{
   \log  \F (c-1, \til{h}_i, \til{h}_{p}, 1-z) \approx -\frac{c}{6} f\Big ( \frac{\tilde{h}_i}{c}, \frac{{\tilde h}_p}{c},1- z \Big ) + \O(c^0),
  }
where $f \approx -({6}/c) \til{h}_p \log (1-z) + \O(1-z)$. In the semiclassical limit the conformal block expansion is dominated by the contribution of light states with small $\tilde{h}_p$. 
In what follows we assume that the vacuum block dominates the conformal block expansion and study its consequences. This is analogous to the assumptions made in the derivation of entanglement entropy on excited states in CFT in refs.~\cite{Caputa:2014vaa,Asplund:2014coa}.\footnote{In sparse CFTs where the density $\rho$ of light states $h + \bar{h} < c/12$ is bounded by $\rho \lesssim e^{\pi c/6}$~\cite{Hartman:2014oaa}, the vacuum block is expected to dominate the conformal block expansion at large $c$~\cite{Hartman:2013mia}. A precise definition of sparseness in WCFTs and its connection to vacuum block dominance have not been spelled out yet. We hope to further understand this assumption and its implications to WCFT holography in the future.}

Note that in WCFTs the vacuum block captures the contribution of all of the descendants of the vacuum including the conserved currents $T(x^+)$ and $P(x^+)$, its derivatives, and its normal-ordered products. The holographic duals of WCFTs are gravitational theories (see Section~\ref{se:intro} for examples) that feature the same Virasoro-Kac-Moody symmetries and the fields dual to the $T(x^+)$ and $P(x^+)$ currents. Therefore, we expect the vacuum block to capture the dynamics of the gravitational sector of the dual theory in the semiclassical limit. For another application of vacuum-block dominance in large-$c$ WCFTs see~\cite{Chen:2018abc}.

We will henceforth approximate the trace of $\rho_{\A}^n$ by
  \eq{
  S_n \approx e^{-i q_n x^{-}_{21}} z^{2 q_{\Psi} q_n/k} (1-z)^{-2 h_n} \F (c-1, \til{h}_i, 0, 1-z).   \label{Ivacuumblock}
  }
The Virasoro vacuum block ${\F} (c-1, \til{h}_i, 0,z)$ is not known in closed form (see e.g.~\cite{Perlmutter:2015iya} for a nice discussion and several series expansions). If we assume that the tilt $\mu$ of the WCFT scales at most linearly with $c$, then the conformal weight and charge of the twist field satisfy
  \eq{
  \frac{\til{h}_n}{c} = \frac{h_n - q_n^2/k}{c} \ll 1, \qquad \textrm{when} \qquad n \to 1.
  }
In this case the semiclassical vacuum block is approximately given by~\cite{Fitzpatrick:2014vua}\footnote{We have implicitly assumed that the descendant states in the vacuum block are unitary. This property does not extend to all blocks in some WCFTs with known holographic duals~\cite{Compere:2009zj,Compere:2013bya}.}
  \eq{
   \F (c-1, \til{h}_i, 0, 1-z) \approx \Bigg [\frac{\g(1-z)z^{\frac{\g -1}{2}}} {1 - z^{\g}} \Bigg ]^{2\til{h}_n}, \label{semiclassicalblock}
  }
where $\g$ is defined by
  \eq{
  \g = \sqrt{1 - 24 \til{h}_{\psi}/c}.
  }
Note that eq.~\eqref{semiclassicalblock} is nonperturbative in $\til{h}_{\psi}/c$ but perturbative in $\til{h}_n/c$. Hence it is valid in the limit $n \to 1$. We conclude that in the semiclassical limit the trace of the reduced density matrix $\rho_{\A}^n$ is given by
  \eq{
 S_{n \to 1} &  \approx e^{-i q_n x^{-}_{21}} z^{2 q_{\psi} q_n /k} (1-z)^{-2 q_n^2/k} \Bigg [\frac{\g z^{\frac{\g -1}{2}}} {1 - z^{\g}} \Bigg ]^{2\til{h}_n}. \label{Infinal}
   }

The entanglement entropy obtained from eq.~\eqref{Infinal} has a nice interpretation when the WCFT is placed on the cylinder. This can be accomplished by the warped conformal transformation
  \eq{
  x^+ = e^{i w^+}, \qquad x^- = w^-,
  }
where $(w^+, w^-) \sim (w^+ + 2\pi, \,w^-)$ are the coordinates on the cylinder. In these coordinates the cross-ratio is given by $z \equiv x^+_1 = e^{i w^+_1}$ and we use $\ell = |w^+_{21}|$ and $\bar{\ell} = |w^-_{21}|$ to denote the lengths of the interval. Thus, the entanglement entropy on the excited state $\psi$ is given by
  %
  \eq{
  S^{\psi}_{\A} &= \lim_{n\to1} \frac{1}{1-n} \log S_n \\
  &= -i P_0^{vac} \bar{\ell} + \mu q_{\psi} \ell -2 \big ( i \mu P_0^{vac} + 2 L_0^{vac} \big ) \log \bigg ( \frac{\til{\b}_{\psi}}{\pi\e} \,\sinh \frac{\pi \ell}{\til{\b}_{\psi}} \bigg ), \label{psiee}
  }
where we used eqs.~\eqref{twistcharges} and~\eqref{vevs}, and $\til{\b}_{\psi}$ is defined by
  \eq{
  \til{\b}_{\psi} = \frac{2\pi}{i \g} = \frac{2\pi}{\sqrt{24 \til{h}_{\psi}/{c} -1}}.
  }

This equation is reminiscent of the entanglement entropy of WCFT at finite temperature~\eqref{thermalee}. This is not a coincidence, as the entanglement entropy on the state $\psi$ can be obtained from eq.~\eqref{psiee} by changing ensembles (canonical $\to$ microcanonical), an observation first made in the context of CFT in~\cite{Asplund:2014coa}. In order to see this let us consider the partition function of a WCFT in the canonical ensemble\footnote{The WCFT partition function is sensitive to the orientation of the spatial circle where the theory is quantized, a consequence of the absence of Lorentz invariance. However, as shown in~\cite{Castro:2015csg}, the expectation values given above turn out to be insensitive to the orientation of the spatial circle.}
  \eq{
  Z(\b, \bar{\b}) = \tr \big( e^{- \b L^{cyl}_0- \bar{\b} P^{cyl}_0} \big). \label{partition}
  }
where $L^{cyl}_0$ and $P^{cyl}_0$ are the zero modes defined on the cylinder. In the high temperature limit $\b \to 0$, the expectation values of $L^{cyl}_0$ and $P^{cyl}_0$ were computed  in~\cite{Castro:2015csg} and read
  \eq{
  \big\langle L_0^{cyl} \big\rangle &= -\frac{\p}{\p\b} Z(\b, \bar{\b}) = \frac{k}{4}\frac{\bar{\b}^2}{\b^2} - 2\pi i P_0^{vac} \frac{\bar{\b}}{\b^2} - 4\pi^2 L_0^{vac} \frac{1}{\b^2}, \label{micro1} \\
    \big\langle P_0^{cyl} \big\rangle &= -\frac{\p}{\p\bar{\b}} Z(\b, \bar{\b}) = -\frac{k}{2} \frac{\bar{\b}}{\b} + 2\pi i P_0^{vac} \frac{1}{\b}. \label{micro2}
  }

Although these expressions were derived in the $\b \to 0$ limit we expect them to be valid for any $\b < 2\pi$ when $c$ is large. This result has been shown to hold in CFTs with a sparse spectrum in~\cite{Hartman:2014oaa} which strongly suggests that a similar result is possible in WCFTs. Using eqs.~\eqref{micro1} and~\eqref{micro2} we find
  \eq{
  -i P_0^{vac}  \bigg ( \frac{\bar{\b}}{\b} - \frac{2\pi\mu}{\b} \bigg ) = \mu q  , \qquad  \b = \frac{2\pi}{\sqrt{24 \til{h}/{c} -1}}, \label{microfinal}
  }
where we used eq.~\eqref{vevs} and defined $\big\langle L_0^{cyl} \big\rangle \equiv h - c/24$ and $ \big\langle P_0^{cyl} \big\rangle \equiv q$. This is the transformation between the canonical and microcanonical ensembles in WCFT. It is not difficult to see that eq.~\eqref{microfinal} with $h = h_{\psi}$ and $q = q_{\psi}$ relates the entanglement entropy at finite temperature~\eqref{thermalee} to the entanglement entropy on the state $\psi$~\eqref{psiee}. 


\section{Local quench} \label{se:quench}

In this section we study the time evolution of $\dd S^{(n)}_{\A}$, the change in the R\'enyi entropy after a local quench. We find that $\dd S^{(n)}_{\A}$ is topological and independent of the size and shape of the subsystem $\A$. While $\dd S^{(n)}_{\A}$ vanishes at early and late times for any WCFT, we show that $\dd S^{(n)}_{\A} \ne 0$ at ``intermediate'' times, but only for charged states in spectrally-flowed WCFTs.


\subsection{R\'enyi entropy via replica}

We begin by specifying what time and space are in the WCFT. As warped CFTs are not relativistic, different choices may lead to different results. In this paper, we align the Virasoro and  $U(1)$ directions with the lightcone coordinates $x^+$ and $x^-$ on the reference plane such that
  \eq{
  x^+ = x + i \tau, \qquad x^ - = x - i \tau, \label{coordinates}
  }
where $\tau$ is the Euclidean time and $x^{\pm}$ transform as in eq.~\eqref{transformation}. This choice is motivated by the identification of coordinates made in known WCFT models~\cite{Compere:2013aya,Hofman:2014loa,Jensen:2017tnb} as well as in holographic duals to WCFTs~\cite{Detournay:2012pc,Compere:2013bya}. An alternative choice corresponds to identifying the Virasoro direction $x^+$ with space and the $U(1)$ direction $x^-$ with time~\cite{Castro:2015csg}. In what follows we find that the dependence on the $U(1)$ coordinate drops out. Hence, in order to probe the real time dynamics of WCFTs the Virasoro direction must depend on time. Another set of coordinates compatible with this requirement identifies the Virasoro direction with time and the $U(1)$ direction with space. This choice is relevant in AdS$_2$ gravity~\cite{Hartman:2008dq,Castro:2008ms,Castro:2014ima,Cvetic:2016eiv,Chaturvedi:2018uov} and the near horizon region of extremal Kerr black holes in four dimensions~\cite{Castro:2009jf,Matsuo:2009sj}. One problem with this identification is the lack of well-defined Fourier expansion and charges. We leave further exploration of these coordinates to future work. 

Let us define the subsystem $\A$ by the interval 
  \eq{
  \A: \quad \big\{ (x,\tau) \big | \,x \in [0, \ell], \tau = 0 \big\}.  \label{interval}
  }
The local quench is modeled by the insertion of an operator $\psi(x,\tau)$ at 
  \eq{
  \Ket{\psi (t)} = \psi(-x_0, -i t ) \Ket{0} = e^{-i t H} \psi(-x_0,0) \Ket{0}, \label{quench}
  }
where $H = - L_{-1} + i P_0$ is the Hamiltonian of the system in the reference plane responsible for the time translation $\p_t$. In a different frame, say $(x^+,y)$, $H$ is given by a different linear combination of the spectrally-flowed $SL(2,R)\times U(1)$ generators, cf.~eq.~\eqref{spectralflowcurrents}. By construction $\psi(x,\tau)$ lies outside the subsystem $\A$ at early ($t=0$) and late times ($t \to \infty$). As time evolves we expect the operator $\psi(x,\tau)$ to cross the interval $\A$.

We are interested in the change of the R\'enyi entropy of subsystem $\A$ before and after the local quench. This quantity is defined as, cf.~eq.~\eqref{renyi},
  \eq{
  \dd S^{(n)}_{\A} = \frac{1}{1-n} \Big \{\log \tr \big [( \rho^{\psi}_\A)^n \big] - \log \tr \big [ (\rho_\A^{\mathbb{1}})^n \big ] \Big \},
  }
where $\rho_{\A}^{\O} = \tr_{\A^{c}} \rho^{\O}$ is the reduced density matrix of subsystem $\A$ on the state $\Ket{\O}$. More concretely, $\rho^{\mathbb{1}} = \Ket{0}\Bra{0}$ denotes the density matrix of the vacuum while $\rho^{\mathbb{\psi}}$ is the density matrix of the state $\Ket{\psi(t)}$ defined by
  \eq{
  \rho^{\psi} = \psi(-x_0,\tau_2) \Ket{0}\Bra{0}\psi^{\dagger}(-x_0,\tau_1), \label{rhopsi}
  }
where $\tau_1$ and $\tau_2$ read
  \eq{
  \tau_1 = \e - it, \qquad \tau_2 = -\e - it. \label{tau12}
  }
We have introduced a regulator $\e$ in eq.~\eqref{tau12} that is necessary to control the UV divergences in the correlation functions of $\psi(x,\tau)$. In particular, when $\e \to 0$ we recover the expected expression for the density matrix on the state $\psi(t)$, namely $\rho = \Ket{\psi(t)}\Bra{\psi(t)}$. 

In this section we evaluate the traces of $(\rho_{\A}^{\psi})^n$ via the replica method,
 \eq{
 \tr \rho_{\A}^n = \frac{Z_n^{\psi}}{\big(Z_1^{\psi}\big)^n}, \label{replica2}
 }
where $Z^{\psi}_n$ is the partition function on $\R_n$ with $2n$ insertions of the operator $\psi$ and $\R_1$ denotes the complex plane. Using eq.~\eqref{replica2} we obtain
  \eq{
  \dd S^{(n)}_{\A} = \frac{1}{1-n} \bigg \{ \log \bigg(\frac{Z^{\psi}_n}{Z^{\mathbb{1}}_n} \bigg ) - n \log  \bigg(\frac{Z^{\psi}_1}{Z^{\mathbb{1}}_{1}} \bigg) \bigg \},
  }
so the change in the $n$-th R\'enyi entropy requires the computation of $2n$-point correlation functions. More concretely, let us denote the coordinates on $\R_n$ by   
  \eq
  {(\til{x}^+, \til{y}) = (\til{x}^+, \til{x}^- + \mu \log \til{x}^+),
  }
where $\til{x}^{\pm} = \til{x} \pm i \til{\tau}$. One then finds
  \eq{
  \frac{Z^{\psi}_n}{Z^{\mathbb{1}}_n} & = \langle \psi^{\dagger}(\til{x}_1, \til{\tau}_1) \psi(\til{x}_2, \til{\tau}_2)\dots \psi^{\dagger}(\til{x}_{2n-1}, \til{\tau}_{2n-1}) \psi(\til{x}_{2n}, \til{\tau}_{2n})\rangle_{\R_n}, \label{npoint}
  }
where the pair of operators $\psi^{\dagger}(\til{x}_{2k-1}, \til{\tau}_{2k-1}) \psi(\til{x}_{2k}, \til{\tau}_{2k})$ is located on the $k^\textrm{th}$ sheet of $\R_n$ such that
  \eq{
arg \big( \til{x}^{\pm}_{2k-1} \big) =  arg \big( x^{\pm}_1 \big) \pm 2\pi i (k-1) , \qquad   arg \big( \til{x}^{\pm}_{2k} \big) =  arg \big(  x^{\pm}_2 \big ) \pm 2\pi i (k-1). \label{Rncoords}
  }
Thus, after an appropriate normalization, $\dd S^{(n)}_\A$ reduces to the $n$-point function of the bilocal operator $\psi^{\dagger}\psi$, namely
  \eq{
  \dd S_{\A}^{(n)} = \frac{1}{1-n} \log \frac{\langle \prod_{k=1}^{n} \psi^{\dagger}(\til{x}_{2k-1}, \til{\tau}_{2k-1}) \psi(\til{x}_{2k}, \til{\tau}_{2k}) \rangle_{\R_n}}{ ( \langle \psi^{\dagger}(\til{x}_1, \til{\tau}_1) \psi(\til{x}_2, \til{\tau}_2)\rangle_{\R_1})^n}. \label{renyin}
  }

We can evaluate the correlation functions on $\R_n$ by exploiting the symmetries of WCFTs. Indeed, eq.~\eqref{correlatorsprime} allows us to express correlators in $\R_n$ in terms of correlators on the complex plane $\C$ provided that a warped conformal transformation exists that maps $\R_n$ to $\C$. This transformation takes the form~\eqref{transformation} and is given by
  \eq{
  z^{+} = \bigg(\frac{\til{x}^+}{\til{x}^+ - \ell} \bigg)^{\frac{1}{n}}, \quad z^{-} = \til{x}^{-} + \mu \bigg( \frac{n-1}{n}\bigg)\log \bigg( \frac{\til{x}^+}{\til{x}^+ - \ell} \bigg) , \label{map}
  }
where $(z^+, z^-)$ denote the coordinates on the reference plane where $\Vev{T(z^+)} = \Vev{P(z^+)} = 0$. A derivation of eq.~\eqref{map} is given in Appendix~\ref{ap:map}. Here we note that eq.~\eqref{map}
\begin{itemize}
 \item[(i)]{ features the desired branch points at $\til{x}^+ = 0$ and $\til{x}^+ = \ell$,}
 \item[(ii)]{ maps the two coordinates $\til{x}^+_{2k-1}$ and $\til{x}^+_{2k}$ on the $k^{\textrm{th}}$ sheet of $\R_n$ to $arg(z^+_{2k-1}) \in \frac{2\pi}{n}(k-\frac{3}{2},k-1)$ and $arg(z^+_{2k}) \in \frac{2\pi}{n}(k-1,k-\frac{1}{2})$,}
 \item[(iii)]{ generalizes the analogous transformation in CFT~\cite{Calabrese:2004eu}.}
\end{itemize}


\subsection{Second R\'enyi entropy after the quench}

We now consider the change in the second R\'enyi entropy $\dd S^{(2)}_{\A}$ in detail and generalize our results to $\dd S^{(n)}_{\A}$ in the next subsection. For $n = 2$, eq.~\eqref{map} maps the $\til{x}^{+}_1$ and $\til{x}^{+}_2$ coordinates to the right half of the complex plane and the $\til{x}^{+}_3$ and $\til{x}^{+}_4$ coordinates to the left half such that
  \eq{
  z^+_3 = e^{\pi i} z^+_1, \qquad z^+_4 = e^{\pi i} z^+_2.
  }
Using eqs.~\eqref{correlatorsprime} and~\eqref{map} the four-point function on $\R_2$ reads
  \eq{
  \!\!\!\! \langle \prod_{k=1}^{2} \psi^{\dagger}(\til{x}_{2k-1}, \til{\tau}_{2k-1}) \psi(\til{x}_{2k}, \til{\tau}_{2k}) \rangle_{\R_2} \! &= \!\bigg [ \frac{\big( (z^+_1)^2-1\big)\big( (z^+_2)^2-1\big)}{2 \ell z^+_{21} }\bigg ]^{4h_{\psi}} \! \frac{ e^{iq_{\psi} (z^-_{21} + z^-_{43})} G(z)}{(z^+_1)^{2h_{\psi}} (z^+_2)^{2h_{\psi}} }, \label{trans4pt}
   }
 where $h_{\psi}$ and $q_{\psi}$ denote the conformal weight and $U(1)$ charge of $\psi$, and the cross-ratio $z$ is now given by
   \eq{
   z = \frac{z_{12} z_{34}}{z_{13}z_{24}}. \label{crossratio2}
   }
Eq.~\eqref{trans4pt} has been written in terms of the $z^-$ coordinate, reason why the spectral flow parameter $\mu$ is not manifest, cf.~the discussion above eq.~\eqref{shift}. The two-point function on $\R_1$ is given by
   \eq{
	 \langle \psi^{\dagger}(\til{x}_1, \til{\tau}_1) \psi(\til{x}_2, \til{\tau}_2)\rangle_{\R_1} & = \frac{e^{i q_{\psi} \til{x}^-_{21}}}{(\til{x}_{12}^+)^{2h_{\psi}}},\label{trans2pt}
  }
and is also independent of $\mu$ in these coordinates. The $\mu$-dependence of the two-point function~\eqref{trans2pt} is manifest when expressed in the same $(z^+, z^-)$ coordinates used in~\eqref{trans4pt}, namely
   \eq{
	 \langle \psi^{\dagger}(\til{x}_1, \til{\tau}_1) \psi(\til{x}_2, \til{\tau}_2)\rangle_{\R_1}  & = \frac{e^{iq_{\psi} [z^-_{21}-\mu \log (z^+_2/z^+_1)]}}{\ell^{2h_{\psi}}} \bigg [ \frac{\big( (z^+_1)^2-1\big)\big( (z^+_2)^2-1\big)}{(z^+_{2})^2 - (z^+_1)^2 }\bigg ]^{2h_{\psi}}. \notag
  }
Thus, the change in the second R\'enyi entropy after the local quench is given by
  \eq{
  \dd S^{(2)}_{\A} = - \log \Big [ e^{iq_{\psi} \mu \log (z^+_2/z^+_1)^2} (1 - z)^{2h_{\psi}} G(z) \Big ],\label{deltas2}
  }
where we have used $z = -(z^{+}_{12})^2/(4z^+_1 z^+_2)$ and $z^-_{43} = z^-_{21}$, equations that follow from the map~\eqref{map}. In particular, in terms of $x_0$, $t$, $\e$, and $\ell$, the cross-ratio satisfies
  \eq{
  z =  \frac{1}{2} \bigg( 1 - \frac{(x_0 - t) (x_0 + \ell - t) + \e^2}{\sqrt{[(x_0-t)^2 + \e^2] [(x_0 + \ell - t)^2 + \e^2]}} \bigg ).
  }
Note that eq.~\eqref{deltas2} is independent of the $U(1)$ direction, i.e.~independent of the $z^-_i$ coordinates. Hence, provided that we set the spectral flow factor to zero, our derivation is also valid for chiral CFTs featuring only one left-moving copy of the Virasoro algebra.\footnote{Although eq.~\eqref{deltas2} with $\mu = 0$ matches the corresponding expression in left-moving chiral CFTs, we expect the chiral CFT function $G(z)$ to differ from that of warped CFT.}

Using the conformal block expansion~\eqref{blockexpansion} we can recast eq.~\eqref{deltas2} as
  \eq{
   \dd S^{(2)}_{\A} = & -\log \Big [ e^{i\mu q_{\psi} \log (z^+_2/z^+_1)^2} (1 - z)^{2\til{h}_{\psi}}   {\textstyle \sum}_{p} (C_{p\psi\psi})^2 \F (c-1,\til{h}_{\psi},\til{h}_p,z) \Big], \label{deltas2blockS}
  }
where we recall that $\til{h}_{\psi}$ and $\til{h}_p$ denote the conformal weights in the Sugawara basis. This expression is valid in the $s$-channel where we expand around $z=0$. Using crossing symmetry, i.e.~by computing the four-point function in the $t$-channel, we obtain an expression that is instead valid around $z=1$. Thus, $\dd S^{(2)}_{\A}$ is alternatively given by
  \eq{
   \dd S^{(2)}_{\A, t- \textrm{ch}} =& -\log \Big [ e^{i \mu q_{\psi} \log (z^+_2/z^+_1)^2} z^{2\til{h}_{\psi}}  {\textstyle \sum}_{p} (C_{p\psi\psi})^2 \F (c-1,\til{h}_{\psi},\til{h}_p, 1-z) \Big].\label{deltas2blockT}
  }

The expansions around the $s$~\eqref{deltas2blockS} and $t$-channels~\eqref{deltas2blockT} capture different regimes in the time evolution of $\dd S^{(2)}_{\A}$.  This follows from the fact that $z$ is a function of time for any value of the regulator $\e$. In particular, for small but non-vanishing $\e$ we have $z \sim \O(\e^2)$ at early and late times whereas $z \sim 1-  \O(\e^2)$ at intermediate times (see below for a precise definition). When the regulator $\e$ is set strictly to zero the cross-ratio becomes
  \eq{
  \lim_{\e \to 0} z = \frac{1}{2} \big \{ 1 - \textrm{sign}[(x_0 -t )(x_0 + \ell -t)] \big \},
  }
and we find more precisely that
  \eq{
  z  =  \left \{ \begin{array}{ll} 0, &\quad 0  \le t < x_0 \textrm{\,\,(early)} \,\, \textrm{and} \,\ x_0 + \ell < t \textrm{\,\,(late)} \\
  1, &\quad x_0 < t < x_0 + \ell \textrm{\,\,(intermediate)}.
  \end{array} \right. \label{crossratiofinal}
  }
In particular we note that $z$ is real, even for finite $\e$, and that it does not circle the point $z = 1$ as the system is evolved in time (instead approaching $z \to 1$ from below). Hence the cross-ratio does not probe the multivaluedness of the four-point function, in contrast to the out-of-time-order correlator considered in the next section. 

For $z = 0$ or  $z = 1$, the leading contribution to each of the conformal blocks $\F$ in eqs.~\eqref{deltas2blockS} and~\eqref{deltas2blockT} is given by the primary state sans descendants, cf. eq.~\eqref{virasoroblock}
  \eq{
  \F (c-1, \til{h}_\psi, \til{h}_{p}, 0) = \lim_{z \to 0} z^{\til{h}_{p}} \big[1 + \O(z)\big].
  }
Since the conformal weight $\til{h}_p$ is invariant under spectral flow, its minimum value is saturated by the vacuum, namely by $\til{h}_p  = 0$. Hence, the conformal block expansion of~\eqref{deltas2} is dominated by the vacuum state in both the $s$ and $t$-channels, and $\dd S^{(2)}_{\A}$ is given by\footnote{When $\e \ne 0$, $z$ varies smoothly as a function of time between $z=0$ and $z=1$. As $z$ grows away from zero the contribution of other primary states to the conformal block expansion becomes important and cannot be neglected. Indeed, as $z$ approaches 1 these states are summed into the conformal block expansion in the $t$-channel which is dominated by the vacuum state at $z = 1$.}
  \eq{
   \dd S^{(2)}_{\A} = & -  i\mu q_{\psi} \log (z^+_2/z^+_1)^2 .\label{deltas22}
  }
Thus, the time-dependence of $\dd S^{(2)}_{\A}$ comes entirely from the $z^+_2/z^+_1$ ratio. Furthermore note that eq.~\eqref{deltas22} implies $\dd S^{(2)}_{\A} = 0$ at all times in a left-moving chiral CFT.
  
The ratio $\xi \equiv (z^+_2/z^+_1)^2$ featured in eq.~\eqref{deltas22} is a pure phase given by
  \eq{
  \xi = e^{i \t} = \frac{(x_0 - t + i\e)(x_0 + \ell - t - i\e)}{(x_0 - t - i\e)(x_0 + \ell - t + i \e)}. \label{xiratio}
  }
 When $\e = 0$, it seems that $\xi = 1$ and $\dd S^{(2)}_{\A}$ vanishes. However, care must be exercised as the limit $\e \to 0$ is taken. Note that as $\epsilon\rightarrow 0^+$ we have the following expansion  
   \eq{
  \xi \sim 1+ { 2i\epsilon \ell\over (x_0-t)(x_0+\ell-t)}, 
  }
For small but finite $\e$, $\t$ evolves counter-clockwise in the complex plane until it reaches $\t = 2\pi$ at $t = x_0$. As time evolves further, $\t$ reverses its path reaching $\t = 0$ after a time $t = x_0 + \ell$.
 Hence, as $\epsilon\to 0^{+}$, the phase takes different values in different regions,
  \eq{
  \lim_{\e\to0^+} \t = \left \{ \begin{array}{ll} 0, &\quad 0  \le t < x_0  \,\, \textrm{and} \,\ x_0 + \ell < t,  \\
  2\pi , &\quad x_0 < t < x_0 + \ell .
  \end{array} \right. \label{theta}
  } 
 As a consequence, the change of the second R\`{e}nyi entropy picks up a phase in the intermediate region such that
  \eq{
  \dd S^{(2)}_{\A} = \left \{ \begin{array}{ll} 0, &\quad 0  \le t < x_0  \,\, \textrm{and} \,\ x_0 + \ell < t,  \\
  2\pi \mu q_{\psi}, &\quad x_0 < t < x_0 + \ell .
  \end{array} \right. \label{deltas2final}
  }
The second R\`{e}nyi entropy is independent of the length $\ell$ of the subsystem. It is in this sense that $\dd S^{(2)}_{\A}$ is topological. Notably, the change in the second R\'enyi entropy after a local quench is nontrivial at intermediate times but only for charged states $\psi$ in spectrally-flowed WCFTs. This is a feature that extends to higher-order R\'enyi entropies as we show next. 
  


\subsection{$n$-th R\'enyi entropy after the quench} 

We now generalize eq.~\eqref{deltas2final} and comment on its interpretation. For arbitrary $n$ the change in the R\'enyi entropy $\dd S^{(n)}_{\A}$ is given, up to a normalization, by a $2n$-point function on $\R_n$~\eqref{renyin}. In particular, this means that we have $n$ pairs of operators $\psi^{\dagger}$ and $\psi$ depending respectively on $n$ pairs of coordinates $\til{x}^{\pm}_{2k-1}$ and $\til{x}^{\pm}_{2k}$ where $k = 1, \dots, n$. The $\til{x}^{\pm}_{2k-1}$ and $\til{x}^{\pm}_{2k}$ coordinates live on the $k$-th sheet of $\R_n$ and are obtained from the $\til{x}^{\pm}_1$ and $\til{x}^{\pm}_2$ coordinates of the original manifold via eq.~\eqref{Rncoords}. As in the previous subsection we will evaluate the $2n$-point function on the complex plane, a feat that is possible due to the properties of the uniformizing map~\eqref{map} that we discuss next.

We begin by noting that to linear order in $\e$, the $z_{2k-1}^+$ and $z^+_{2k}$ coordinates obtained via eqs.~\eqref{Rncoords} and~\eqref{map} satisfy
  \eqsp{
 \big  (z^+_{2k-1},z^+_{2k} \big) =  \left \{ \begin{array}{ll} 
 e^{\frac{2\pi i (k-1)}{n}}\zeta \big( e^{-\frac{i \d}{n}}, e^{\frac{i \d}{n}} \big) , & \quad 0  \le t < x_0  \,\, \textrm{and} \,\ x_0 + \ell < t, \\
 e^{\frac{2\pi i (k-1)}{n}} \zeta \big( e^{-\frac{i\pi}{n} - \frac{i\d}{n}}, e^{\frac{i\pi}{n} + \frac{i\d}{n}} \big), & \quad x_0  \le t < x_0 + \ell, 
   \end{array} \right. \label{zcoords}
   }
where $\zeta$ and $\d$ are defined as
  \eq{
  \zeta \equiv \Big | \frac{x_0 - t}{x_0 + \ell - t} \Big |^{\frac{1}{n}} , \qquad \d = \frac{ \ell \e}{(x_0 - t)(x_0 +\ell - t)}.
  }
In particular, eq.\eqref{zcoords} implies that the coordinates pair up at early and late times as
  \eq{
 & z^+_1 z^+_2 \dots z^+_{2n-1} z^+_{2n} \to (z^+_1 z^+_2)\dots (z^+_{2k-1} z^+_{2k}) \dots ( z^+_{2n-1} z^+_{2n}), \label{earlylate2n}
 }
where the notation $(z^+_i z^+_j)$ means that $z^+_i - z^+_j \propto \e$. As $\e\to 0$ there is a sharp transition at intermediate times where we instead find
 \eq{
 & z^+_1 z^+_2 \dots z^+_{2n-1} z^+_{2n} \to (z^+_2 z^+_3)\dots (z^+_{2k} z^+_{2k+1}) \dots ( z^+_{2n} z^+_{1}). \label{intermediate2n}
  }
When $n =2$ eqs.~\eqref{earlylate2n} and~\eqref{intermediate2n} are compatible with the behavior of the cross-ratio found in the previous subsection. There, $z \to 0$ at early and late times while $z \to 1$ at intermediate times~\eqref{crossratiofinal}.
  
The $2n$-point function featured in $\dd S^{(n)}_{\A}$ is given via the uniformizing map~\eqref{map} by
  \eq{
 \langle \prod_{k=1}^{n} \psi^{\dagger}(\til{x}_{2k-1}, \til{\tau}_{2k-1}) \psi(\til{x}_{2k}, \til{\tau}_{2k}) \rangle_{\R_n} & = \bigg ( \frac{ \zeta \d}{n \e} \bigg )^{2 n h_{\psi}} \langle \prod_{k=1}^{n} \psi^{\dagger}(z^+_{2k-1}, z^-_{2k-1}) \psi(z^+_{2k},z^-_{2k}) \rangle.
   }
As a consequence of eqs.~\eqref{earlylate2n} and~\eqref{intermediate2n} the $n$-point function of the bilocal operator $\psi^{\dagger}(z^+_{2k-1},z^-_{2k-1})\psi(z^+_{2k},z^-_{2k})$ on the complex plane factorizes accordingly into the product of $n$ two-point functions. Interestingly, we find that
  \eq{
 \langle \prod_{k=1}^{n} \psi^{\dagger}(\til{x}_{2k-1}, \til{\tau}_{2k-1}) \psi(\til{x}_{2k}, \til{\tau}_{2k}) \rangle_{\R_n}  & = \frac{\prod_{k=1}^n e^{i q_{\psi} ( z^-_{2k} - z^-_{2k-1} )} }{(2i \e)^{2 n h_{\psi}}}, \label{2ndeltasn}
  }
is valid at all times. To obtain the change in the R\'enyi entropy $\dd S^{(n)}_{\A}$ we must first normalize eq.~\eqref{2ndeltasn} by $n$ copies of the two-point function
  \eq{
 \langle \prod_{k=1}^{2} \psi^{\dagger}(\til{x}_{2k-1}, \til{\tau}_{2k-1}) \psi(\til{x}_{2k}, \til{\tau}_{2k}) \rangle_{\R_1} & = \frac{e^{i q_{\psi} \til{x}^-_{21}}}{(2ix)^{2h_{\psi}}}.
  }
Thus, since $\til{x}_{21} = z^-_{21} + \mu (1-n) \log (z^+_2/z^+_1)$ and $z^-_{2k} - z^-_{2k-1} = z^-_{21}$ for all $k$, the change in the $n$-th R\'enyi entropy is given by
  \eq{
   \dd S^{(n)}_{\A} =  - i \mu q_{\psi} \log (z^+_2/z^+_1)^n. \label{deltasn}
  }

When $n=2$ in eq.~\eqref{deltasn} we recover the results found in the previous subsection. One motivation behind the more detailed analysis of the $n =2$ case concerns the analytic properties of the four-point function. Indeed, when $n =2$ we observed that the cross-ratio does not circle the branch point at $z = 1$. Instead, a nontrivial value of $\dd S^{(2)}_{\A}$ comes solely from the multivaluedness of the $U(1)$ phases in the two and four-point functions of WCFTs. Motivated by the $n=2$ result, we have implicitly assumed that the multivaluedness of $\dd S^{(n)}_{\A}$ derived in eq.~\eqref{deltasn} comes only from the $U(1)$ phases. Thus, provided this assumption holds, the change in the $n$-th R\'enyi entropy is given by
  \eq{
  \dd S^{(n)}_{\A} = \left \{ \begin{array}{ll} 0, &\quad 0  \le t < x_0  \,\, \textrm{and} \,\ x_0 + \ell < t,  \\
  2\pi \mu q_{\psi}, &\quad x_0 < t < x_0 + \ell .
  \end{array} \right. \label{deltasnfinal}
  }
where we used the properties of the ratio $(z^+_2/z^+_1)^n = \xi = e^{i\t}$ studied in the previous subsection, namely the fact that $\t = 2\pi$ at intermediate times and vanishes otherwise~\eqref{theta}.

Eq.~\eqref{deltasnfinal} highlights the importance of spectral flow in the study of WCFTs, the latter of which is necessary to obtain nontrivial values of $\dd S^{(n)}_{\A}$. Furthermore, eq.~\eqref{deltasnfinal} vanishes unless the operator $\psi$ is charged, i.e.~unless it carries momentum along $x^-$. In fact, the only non-vanishing contribution to $\dd S^{(n)}_{\A}$ comes from the $U(1)$ sector of the WCFT. Thus, eq.~\eqref{deltasnfinal} further distinguishes warped from chiral CFTs, as internal $U(1)$ symmetries do not lead to the spacetime-dependent phases in correlation functions that are needed to obtain a nonvanishing $\dd S^{(n)}_{\A}$.

In refs.~\cite{Nozaki:2014hna,Nozaki:2014uaa,He:2014mwa}, eq.~\eqref{deltas2final} was argued to be compatible with the causal structure of relativistic field theories. While WCFTs are inherently nonrelativistic and nonlocal, our results suggest a similar interpretation. Namely, that the operator $\psi(-x_0,0)$ creates an entangled pair of particles at $x=-x_0$ that propagate oppositely at the speed of light. This is compatible with the fact that $\dd S^{(n)}_{\A} \ne 0$ at $x_0 < t < x_0 + \ell$, that is, when one of the particles lies in subsystem $\A$. It would be interesting to relate our results to the observation that the causal domain of WCFT is a strip~\cite{Castro:2015csg}, as opposed to a diamond, and to understand the consequences to the causal structure of the holographic dual theories (see~\cite{Wen:2018mev} for a related discussion). It would also be interesting to understand the physical interpretation behind the $\mu$-dependence of $\dd S^{(n)}_{\A}$.


\subsection{Comments on different approaches}

We conclude this section with a few comments on different approaches to the calculation of the entanglement entropy after a local quench, specifically refs.~\cite{Asplund:2014coa} and~\cite{He:2014mwa}. In these references the local quench is also modeled by the insertion of a local operator $\psi$. However, while ref.~\cite{Asplund:2014coa} uses the four-point function with twist operators~\eqref{twist4pt} to evaluate the entanglement entropy, ref.~\cite{He:2014mwa} uses the replica four-point function~\eqref{trans4pt} instead. For this reason, the entanglement entropy computed in these approaches exploits the nonanalytic structure of CFT $n$-point functions in different ways. 

In more detail, ref.~\cite{Asplund:2014coa} computed the entanglement entropy after a local quench in large-$c$ CFTs in the semiclassical limit. Therein the vacuum block is assumed to dominate in each channel of the conformal block expansion and different OPE channels dominate at different times. In this case a nontrivial value for the entanglement entropy is obtained at intermediate times when the cross-ratio in the right-moving sector, namely $\bar{z}$, moves onto the second sheet of the vacuum block. On the other hand, ref.~\cite{He:2014mwa} considered the change in the $n$-th R\'enyi entropy of rational CFTs. There, instead of exploiting the analytic properties of $2n$-point functions, the authors use fusion rules to relate (the chiral half of) the vacuum block in one OPE channel to the sum over blocks in different OPE channels. This procedure yields nontrivial values for $\dd S^{(n)}_{\A}$ at intermediate times only.

The change in the R\'enyi entropy of WCFTs computed in  this paper differs from the previous approaches in a novel way. First of all, due to the chirality of the symmetry algebra, the correlation functions can be readily evaluated at all times when $\e \to 0$, and the resulting $2n$-point functions are multivalued. This feature is used at intermediate times when the argument of the correlation function moves onto a different branch. However, unlike ref.~\cite{Asplund:2014coa}, the nonanalyticity of the four-point function originates from the $U(1)$ sector and is present only in spectrally-flowed WCFTs. As a consequence, $\dd S^{(n)}_{\A} \ne 0$ at intermediate times provided that the tilt $\mu$ of the theory and the $U(1)$ charge $q_{\psi}$ of the state $\psi$ do not vanish.


\section{Butterflies in WCFT}  \label{se:chaos}

In this section we show that large-$c$ WCFTs are maximally chaotic provided that the vacuum block dominates the conformal block expansion of the four-point function. As in Section~\ref{se:excited} we also consider operators whose conformal weights scale with the central charge. These set of assumptions are holographic in nature and the maximal Lyapunov exponent obtained herein suggests the corresponding WCFTs are dual to theories of gravity that admit black hole solutions.


\subsection{Semiclassical limit and the vacuum block}

In order to probe the chaotic behavior of WCFTs we turn our attention to the out-of-time order four-point function of two generic operators $W(t)$ and $V(0)$,
  \eq{
  R(t) = \frac{\Vev{W^{\dagger}(t) V^{\dagger}(0) W(t) V(0)}_{\b}}{\Vev{W^{\dagger}(t) W(t)}_{\b} \Vev{V^{\dagger}(0) V(0)}_{\b}}, \label{oto}
  }
where $t$ is the Lorentzian time. In chaotic systems the OTO correlator~\eqref{oto} decays exponentially at late times, a phenomenon that is absent in time-ordered correlators~\cite{Maldacena:2015waa}. The late-time behavior of~\eqref{oto} is preceded by a rapid decrease, $R(t) \approx 1 - f_0 e^{\l_L t}$ with $f_0 \ll 1$, that is characterized by the Lyapunov exponent $\l_L$. The latter is bounded by $\l_L \le 2\pi/\b$ in systems with a parametrically large number of degrees of freedom (large $c$)~\cite{Maldacena:2015waa}. In particular, the Lyapunov exponent vanishes in theories without chaos, e.g.~integrable models, chiral CFTs~\cite{Perlmutter:2016pkf}, and permutation orbifolds~\cite{Caputa:2017rkm,Belin:2017jli}, while its upper bound is saturated by black holes in Einstein gravity. For this reason, we expect theories with $\l_L = 2\pi/\b$ to admit gravitational duals in the large-$c$ limit~\cite{Shenker:2013pqa,Shenker:2013yza,Shenker:2014cwa,Leichenauer:2014nxa}.

We begin by considering the four-point function~\eqref{oto} in Euclidean time. This finite-temperature correlator can be obtained from the four-point function in the complex plane via the change of coordinates~\eqref{mapbeta}, reproduced here for convenience
  \eq{
  x^+ = e^{\frac{2\pi}{\b} w^+}, \qquad x^- = w^- + \bigg ( \frac{\bar{\b}}{\b} - \frac{2\pi\mu}{\b} \bigg ) w^+,
  }
where ${w}^{\pm} = \none{x} \pm i \none{\tau}$ denote the coordinates on the ``thermal" cylinder. The equation above features a chemical potential $\bar{\beta}$ for the $U(1)$ charge and a nonvanishing tilt $\mu$. However, unlike Sections~\ref{se:excited} and~\ref{se:quench}, neither $\bar{\beta}$ nor $\mu$ play a role in the OTO correlator  in our setup. Using eq.~\eqref{correlatorsprime} we find
  \eq{
  R(z) &\equiv  \frac{\big \langle W^{\dagger}(\none{x}_1,\none{\tau}_1) W (\none{x}_2,\none{\tau}_2) V^{\dagger}(\none{x}_3, \none{\tau}_3) V(\none{x}_4, \none{\tau}_4) \big\rangle_{\b}} {\big\langle{W^{\dagger}(\none{x}_1, \none{\tau}_1) W(\none{x}_2, \none{\tau}_2)}\big\rangle_{\b} \big\langle{V^{\dagger}(\none{x}_3, \none{\tau}_3) V(\none{x}_4, \none{\tau}_4)}\big\rangle_{\b}} = G(z), \label{oto2}
  }
where we have assumed the operators $W(\none{x}_i,\none{\tau}_i)$ and $V(\none{x}_j,\none{\tau}_j)$ have unit norm and $z$ is the cross-ratio
  \eq{
  z = \frac{x^+_{12} x^+_{34}}{x^+_{13} x^+_{24}}.
  }

We note that eq.~\eqref{oto2} does not depend on $w^-_i$ and is a multivalued function of $z$. In contrast to correlation functions in CFT, the warped CFT four-point function is multivalued even before analytic continuation to Lorentzian time. This is a consequence of the chiral nature of the Virasoro-Kac-Moody algebra. Aside from the $U(1)$ phases, the correlation functions in WCFT correspond to the chiral half of CFT. This partly motivates using the Euclidean theory as our starting point, since several results, including the conformal block expansion and the approximate expressions for individual blocks and their properties are already known from Euclidean CFT.

If we assume the Virasoro-Kac-Moody vacuum block dominates the conformal block expansion of eq.~\eqref{oto2} we obtain, cf.~eq.~\eqref{block},
  \eq{
  R(z) \approx (1-z)^{{-2q_W q_V/k}} {\F} (c-1, \til{h}_i, 0,z), \label{otovac}
  }
where $\til{h}_i = \{ \til{h}_W, \til{h}_V\}$ are the conformal weights of $W$ and $V$ in the Sugawara basis, and $q_W$, $q_V$ are their $U(1)$ charges. Furthermore, in the semiclassical limit\footnote{Note that a small $\til{h}_W/c$ is a further assumption beyond the semiclassical limit considered in Section~\ref{se:excited}.}
  \eq{
  c \to \infty, \qquad \frac{\til{h}_W}{c}, \frac{\til{h}_V}{c} \ll 1,
  \label{semiclassical}
  }
the expression for the vacuum block given in eq.~\eqref{semiclassicalblock} yields
  \eq{
   R(z) \approx (1-z)^{\g} \bigg [\frac{z}{1 - (1-z)^{1 - 12 \til{h}_W/c}} \bigg ]^{2\til{h}_V}, \label{otomulti}
  }
where $\g$ is given by
  \eq{
  \g = - \frac{12\til{h}_W \til{h}_V}{c} -  \frac{2q_W q_V}{k}. \label{gamma}
  }
The branch cut extending along $z = [1, \infty)$ is a crucial feature of eq.~\eqref{otomulti}.
It is ultimately responsible for the chaotic behavior of the OTO correlator to be derived in eq.~\eqref{otofinalt}. 
The same nonanalytic behavior~\eqref{otomulti} appears in both the holomorphic and antiholomorphic sectors of CFT~\cite{Roberts:2014ifa} but the full Euclidean correlator remains analytic. In this case the analytic continuation to Lorentzian time breaks the symmetry between the two chiral halves of the four-point function and different time orderings explore different branches of one of the chiral sectors. In contrast, WCFTs are already chiral but an appropriate time ordering prescription is required to probe the nonanalytic behavior of~\eqref{otomulti}, as we will show in the next subsection.

Before we consider the OTO correlator let us comment on chiral CFTs which also feature a chiral symmetry algebra.\footnote{Chiral CFTs could have different meanings in different contexts. See~\cite{Chaturvedi:2018uov} for a partial classification.} In chiral CFTs with $c=24n$ and $n \in \mathbb{N}$~\cite{Maloney:2009ck}, all primary states are conserved currents with integer conformal weights. Consequently, eq.~\eqref{oto2} is single-valued and features no chaos~\cite{Perlmutter:2016pkf}. This is not expected to be the case in WCFT where the conformal weights are not assumed to be integers except for the states corresponding to the vacuum, the $U(1)$ current $P(x^+)$, and their Virasoro descendants. In particular, the analyticity of the four-point function in chiral CFTs implies that the vacuum block cannot be assumed to dominate the conformal block expansion at large $c$. In this case, additional blocks must be included which restore single-valuedness of the four-point function. From the point of view of the conformal block expansion this is possible due to fine-tuning of the OPE coefficients and the conformal weights. Thus, in the derivation of eq.~\eqref{otomulti} we have assumed that WCFTs feature no such fine tuning. 
  

\subsection{Out-of-time order correlator and chaos}

Although eq.~\eqref{otomulti} is multivalued, not all orderings of the operators $W(\none{x}_i,\none{\tau}_i)$ and $V(\none{x}_j,\none{\tau}_j)$ probe different sheets in real time. When eq.~\eqref{oto2} is confined to the first sheet, the correlator remains of $\O(1)$ at late times. Only out-of-time-order correlators probe the second sheet and decay exponentially at late times~\cite{Roberts:2014ifa}. This motivates the following choice of coordinates for the operators $W(\none{x}_i,\none{\tau}_i)$ and $V(\none{x}_j,\none{\tau}_j)$ in eq.~\eqref{oto2}
  \eq{
  {w}^+_k = \left \{ \begin{array}{lcl} -x_0 + i (\e_k - i t), &&\quad k = 1, 2, \\
  i \e_k, && \quad k = 3, 4,
  \end{array} \right. \label{positions}
  }
where $x_0 > 0$ and the $\e_i$ terms are small but finite and chosen to satisfy\footnote{The coordinates ${w}^-_i$ do not play a role in our story and can be obtained from conjugation of ${w}^+_i$ if $\e_k - it$ is assumed to be real. }
  \eq{
  \e_1 > \e_3 > \e_2 > \e_4. \label{epsilons}
  }
The Euclidean-time ordering of the operators $W(\none{x}_i,\none{\tau}_i)$ and $V(\none{x}_j,\none{\tau}_j)$ given by eq.~\eqref{epsilons}  corresponds to the desired out-of-time ordering of the correlator $R(t)$ in eq.~\eqref{oto}. This choice of coordinates implies that the cross-ratio $z$ evolves from $z=0$ at early times to
  \eq{
  z \approx - e^{\frac{2\pi}{\b} (x_0 - t) } \e_{12}^* \e_{34}, \qquad \textrm{at } t \gg x_0, \label{crossratiolatetime}
  }
where we have defined $ \e_{jk}$ as in~\cite{Roberts:2014ifa} by
 \eq{
 \e_{jk} = i \big ( e^{\frac{2\pi}{\b} i \e_j} - e^{\frac{2\pi}{\b} i \e_k} \big ).
 }

The ordering~\eqref{epsilons} guarantees that $z$ evolves counterclockwise in the complex plane circling once the branch point at $z = 1$. As a result, the late time behavior of the OTO correlator is obtained by taking the limit $z \to 0$ on the second sheet of eq.~\eqref{otomulti}. Using the late time expression of the cross-ratio~\eqref{crossratiolatetime} we find
  \eq{
  R(t) \approx e^{-2\pi i \g} \Bigg [ \frac{1}{1 + \frac{24\pi i \til{h}_W}{\e^{*}_{12} \e_{34} }e^{\frac{2\pi}{\b} (t - t_{\star} -x_0) }}
  \Bigg ]^{2 \til{h}_V}, \label{otofinalt}
  }
where we defined the scrambling time $t_{\star}$ by~\cite{Roberts:2014ifa} (see also~\cite{Sekino:2008he,Lashkari:2011yi})
  \eq{
  t_{\star} = \frac{\b}{2\pi} \log c.
  }
The OTO correlator~\eqref{otofinalt} decreases by a term proportional to $- e^{\frac{2\pi}{\b} (t - t_{\star} - x_0 )}$ at times $\b/2\pi < t < t_{\star} + x_0$. This suggests that large-$c$ WCFTs where the vacuum block dominates the conformal block expansion are maximally chaotic and saturate the bound on the Lyapunov exponent
  \eq{
  \l_L = \frac{2\pi}{\b}.\label{lyaponov}
  }
We therefore expect WCFTs satisfying the aforementioned assumptions to admit gravitational duals featuring black holes. On the other hand, the OTO correlator decays as $e^{-{4\pi \til{h}_V}/{\b} (t - t_{\star} - x_0)}$ at late times $ t > t_{\star} + x_0$, which is the expected behavior in chaotic systems.

Up to the phase factor, eq.~\eqref{otofinalt} agrees with the CFT out-of-time-order correlator after the identification $\til{h}_i \to h_i$~\cite{Roberts:2014ifa}. From a technical point of view, the main difference between the warped and CFT calculations comes from the expansion of the four-point function in terms of Virasoro-Kac-Moody blocks. This leads to the conformal weights in the Sugawara basis featured in eq.~\eqref{otofinalt}. Physically, the similarities between the warped and CFT calculations are to be expected from a generalization of the arguments given in ref.~\cite{Perlmutter:2016pkf}. Therein it was argued that the leading contribution to the OTO correlator in a holographic CFT comes from the current of highest spin, namely the stress energy tensor. In contrast to CFTs with Virasoro or $\W_N$ algebras, warped CFTs feature an additional spin-1 current $P(x^+)$ which, as we have seen, does not spoil the leading order contribution of the spin-2 current $T(x^+)$.

Holographic duals for WCFTs come in a variety of flavors, including topologically massive gravity~\cite{Compere:2009zj}, stringy truncations that include matter~\cite{Detournay:2012dz}, Einstein gravity with modified boundary conditions~\cite{Compere:2013bya}, and lower-spin gravity~\cite{Hofman:2014loa}. It is not clear which of these theories is compatible with vacuum-dominance in the dual large-$c$ WCFT. It would be interesting to use the shock wave methods of refs.~\cite{Shenker:2013pqa,Shenker:2013yza,Roberts:2014isa} to answer this question and reproduce eq.~\eqref{otofinalt} directly from gravity.

Finally note that the OTO correlator in CFTs featuring an internal $U(1)$ symmetry, semiclassical limit~\eqref{semiclassical}, and vacuum block dominance, is also given by eq.~\eqref{otofinalt}. This follows from the fact that (i) a $U(1)$ current enhances the Virasoro algebra of CFT to the Virasoro-Kac-Moody algebra of WCFT, and (ii) features characteristic of WCFTs such as spectral flow do not play a role in the OTO four-point function~\eqref{otofinalt}.





\section*{Acknowledgments}
We are grateful to Alex Belin, Pankaj Chaturvedi, Bin Chen, Peng-xiang Hao, Max Riegler, Boyang Yu, and Yuan Zhong for helpful discussions. This work was supported by the National Thousand-Young-Talents Program of China and NFSC Grant No.~11735001. The work of L. \!A. was also supported by the International Postdoc Program at Tsinghua University. S. \!H. is supported by a Max-Planck Fellowship and the German-Israeli Foundation for Scientific Research and Development.


\appendix

\section{The Virasoro-Kac-Moody block} \label{ap:block}

In this appendix we derive the Virasoro-Kac-Moody block given in eq.~\eqref{block}. A convenient way to compute conformal blocks is to insert the identity operator $\sum_{p}\Ket{\til{\a}_p}\Bra{\til{\a}_p}$ in the four-point function~\eqref{4pt}, where the sum runs over primary states with conformal weight $h_p$ and charge $q_p = q_1 + q_2 = - (q_3 + q_4)$ (the $q_i$ charges, where $i = 1, \dots, 4$, are defined below). Using eqs.~\eqref{4pt} and~\eqref{blockexpansion} we have
 \eq{
 \V_{p}(z) =  \lim_{x_1^+ \ra \infty} (x^+_1)^{2 h_1} z^{h_3 + h_4} {e^{-{i} \sum_{j}q_{j} x^-_{j}}} \frac{\Vev{\phi_1(x_1^+) \phi_2(1) \Ket{\til{\a}_{p}} \Bra{\til{\a}_{p}} \phi_3(z) \phi_4(0)}}{C_{p12}C^p{}_{34}},   \label{vkmblock}
 }
 where $z$ is the cross-ratio~\eqref{crossratio}, the operators $\phi_i(x^+_i)$ have weight $h_i$ and charge $q_i$, and we have suppressed their $x^-_i$-dependence for convenience. The $\Ket{\til{\a}_{p}}\Bra{\til{\a}_{p}}$ term in eq.~\eqref{vkmblock} sums the contributions of all the descendant states of the primary state $\Ket{h_p, q_p}$. In the Sugawara basis it reads
  \eq{
  \Ket{\tilde{\a}_{p}}\Bra{\tilde{\a}_{p}} = \sum_{N,M} \sum_{\{n_i, \rho_i\}}  \sum_{\{m_j, \rho_j\}} \big | {\til{\a}_{p}^{N_i,M_j}} \big\rangle \big\langle{\til{\a}_{p}^{N_i,Mj}}\big|,
  }
where $N_i$ and $M_i$ are shorthand for $N_i \equiv \{N, \{n_i, \rho_i\}\}$, $M_j \equiv \{M, \{m_j, \s_j\}\}$ and $\big | {\til{\a}_{p}^{N_i,M_j}} \big\rangle$ is given by
  \eq{
  &\big | {\til{\a}_{p}^{N_i,M_j}} \big\rangle = \bigg [ \tilde{\N}_{\{n_i, \rho_i \}}^{-1/2}  \prod_{i=0}^{N} \big (\tilde{L}_{-n_i} \big)^{\rho_i} \bigg ]\bigg [ \K_{\{m_j, \s_j \}}^{-1/2}  \prod_{j=0}^{M} \(P_{-m_j} \)^{\s_j} \bigg ] \Ket{h_p, q_p},   \label{virkmdescendants}
  }
In eq.~\eqref{virkmdescendants} $\big|{\tilde{\a}_{p}^{N_i,M_j}}\big\rangle$ denotes the state with $\sum_i \rho_i = N$ Virasoro and $\sum_j \s_j = M$ $U(1)$ descendants, $\til{\N}_{\{n_i,\rho_i\}}\times \K_{\{m_j,\s_j\}}$ is the norm of the state, and $\tilde{L}_n$ denotes the Virasoro modes in the Sugawara basis, namely
  \eq{
  \tilde{L}_{n} = L_{n} - \frac{1}{k} \sum_{m \le -1} P_m P_{n-m} + \sum_{m \ge 0} P_{n-m}P_m. \label{sugawaramodes}
  }
In the Sugawara basis the symmetry algebra simplifies to the direct product of a Virasoro and $U(1)$ algebras with $c \ra c - 1$. As a consequence, the Virasoro and $U(1)$ descendant states are orthogonal to each other --- reason why the norm factorizes in eq.~\eqref{virkmdescendants} --- and the conformal block $\V_p(z)$ splits into the product of Virasoro and $U(1)$ Kac-Moody blocks,
  \eq{
  \V_{p}(z) = \V^{Vir}_{p}(z)\V^{KM}_{p}(z).\label{productblock}
  }

The derivation of eq.~\eqref{productblock} proceeds as follows~\cite{Fitzpatrick:2015zha}. We first consider the contribution of one Virasoro descendant to $\Ket{\tilde{\a}_{p}}\Bra{\tilde{\a}_{p}}$, i.e.~the contribution of the state $\big|{\tilde{\a}_{p}^{N_1,0}}\big\rangle  = \wtil{\N}_{\{n,1\}}^{-1/2} \tilde{L}_{-n} \Ket{h_p, q_p}$ where $N_1 = \{1, \{n,1\}\}$. Let $\phi_i \equiv \phi_i(x^+_i, x^-_i)$. Using
  \eq{
  [L_n, \phi_i ] &= \big [ h_i (n+1) (x^+_i)^n + (x^+_i)^{n+1} \p_{x^+_i} \big ] \phi_i, \qquad [P_n, \phi_i] = q_i (x^+_i)^n \phi_i, \label{transformationprimaries}
  }
and eq.~\eqref{sugawaramodes}, it is straightforward to show that
  \eqsp{
 \big \langle {\til{\a}_{p}^{N_1,0}} \big | {\phi_1 \phi_2} \big \rangle =  \,\,& \wtil{\N}_{\{n,1\}}^{-1/2}\bigg \{ {\textstyle \sum}_{i=1}^2 \Big [(n+1) \tilde{h}_i  \,(x^+_i)^n  + (x^+_i)^{n+1} \p_{x^+_i} \Big ]  \\
& \hspace{3cm}-\frac{2 q_1 q_2}{k}  \cdot\frac{(x_1^+)^{n+1} - (x^+_2)^{n+1}}{x^+_1 - x^+_2} \bigg \} \! \BraKet{h_p, q_p}{\phi_1 \phi_2}{}, \label{step2}
  }
where $\til{h}_i = h_i - q_i^2/k$ (the eigenvalue of $\til{L}_0$) is the conformal weight in the Sugawara basis. Crucially, the three-point function in eq.~\eqref{step2} may be written as
  \eq{
  \BraKet{h_p, q_p}{\phi_3 \phi_4}{} = C_{p34}\frac{e^{\frac{i}{3} (q_3 x^-_3 + q_4 x^-_4)}}{(x^+_{34})^{h_3 + h_4 - h_p}} = (x^+_{34})^{\frac{2 q_3 q_4}{k}}   \BraKet{h_p, q_p}{\phi_3 \phi_4}{}_{h_i \ra \tilde{h}_i, \,h_p \ra \tilde{h}_p}, \label{3ptshift}
  }
where we used $q_p = q_3 + q_4$. When the factor of $(x^+_{34})^{{2 q_3 q_4}/{k}}$ is moved through the derivatives in eq.~\eqref{step2}, the three-point function can be expressed in terms of $\tilde{h}_i$ and $\tilde{h}_p$ such that
  \eq{
  \big \langle {\til{\a}_{p}^{N_1,0}} \big | {\phi_3\phi_4} \big \rangle  & = (x^+_{34})^{\frac{2 q_3 q_4}{k}}  \big \langle {{\a}_{p}^{N_1}} \big | {\phi_3\phi_4} \big \rangle{}_{h_i \ra \tilde{h}_i, \,h_p \ra \tilde{h}_p, \,c \ra c -1}, \label{onevirdesc}
  }
where $\big | {\a_{p}^{N_i}} \big\rangle$ denotes the normalized state with $\sum_i \rho_i = N$ Virasoro descendants, namely\footnote{The shift in the central charge in  eq.~\eqref{onevirdesc} is necessary to reproduce the correct normalization, that is $\wtil{\N}_{\{n_i,\rho_i\}} = \N_{\{n_i,\rho_i\}}\Big|_{h_p \ra \tilde{h}_p, \,c \ra c -1}$.}
  \eq{
 \big | {\a_{p}^{N_i}} \big\rangle  = \N_{\{n_i, \rho_i \}}^{-1/2}  \prod_{i=1}^{N} \big ({L}_{-n_i} \big)^{\rho_i}\Ket{h_p, q_p}. \label{virasorodescendants} 
  }

The commutativity of the $\tilde{L}_{n}$ and $P_m$ generators allows us to generalize eq.~\eqref{onevirdesc} to states with an arbitrary number of $\til{L}_n$ descendants. There, the role of the $(x^+_{34})^{2 q_3 q_4/k}$ term in~\eqref{3ptshift} is to give, upon differentiation, the appropriate powers of the charges $q_i$ such that $h_i \ra \tilde{h}_i$. Thus, more generally we find that
  \eq{
  \big \langle {\til{\a}_{p}^{N_i,0}} \big | {\phi_3 \phi_4} \big \rangle & =  (x^+_{34})^{\frac{2 q_3 q_4}{k}}  \big \langle {{\a}_{p}^{N_i}} \big | {\phi_3 \phi_4} \big \rangle{}_{h_i \ra \tilde{h}_i, \,h_p \ra \tilde{h}_p, \,c \ra c -1}. \label{part1}
  }
Similarly, by first considering the case with one Virasoro descendant state and then exploiting the virtues of the Sugawara basis we obtain
  \eq{
\big \langle { {\phi_1 \phi_2}} \big | \til{\a}_{p}^{N_i,0} \big \rangle & = \frac{(x^+_{12})^{\frac{2q_1 q_2}{k}}}{ (x^+_1)^{\frac{2 q_1(q_1 + q_2)}{k}} (x^+_2)^{\frac{2 q_2 (q_1 + q_2)}{k}}} \big \langle { {\phi_1\phi_2}} \big | \a_{p}^{N_i} \big \rangle{}_{h_i \ra \tilde{h}_i, \,h_p \ra \tilde{h}_p, \, c \ra c-1}. \label{part2}
  }
Finally, using eqs.~\eqref{part1} and~\eqref{part2} we find that $\V^{Vir}_{p}(z)$ in~\eqref{productblock} is given by
  \eq{
  \V^{Vir}_{p}(z) &= \lim_{x^+_1 \ra \infty} (x^+_1)^{2 h_1} z^{h_3 + h_4} {e^{-{i} \sum_{j}q_{j} x^-_{j}}} \sum_{N} \sum_{\{n_i, \rho_i\}}   \frac{\big \langle \phi_1 \phi_2 \big | {\til{\a}_{p}^{N_i,0}}\big \rangle \big \langle {\til{\a}_{p}^{N_i,0}} \big | \phi_3 \phi_4 \big \rangle}{C_{p12}C^p{}_{34}}, \notag \\
  & =  {z^{\frac{(q_3 + q_4)^2}{k}}}{}\! \lim_{x^+_1 \ra \infty} \Big [ (x^+_1)^{2 {h}_1} z^{{h}_3 + {h}_4} {e^{-{i} \sum_{j}q_{j} x^-_{j}}} \sum_{N} \!\!\sum_{\{n_i, \rho_i\}} \!\! \frac{\big \langle \phi_1 \phi_2 \big | {\a_{p}^{N_i}}\big \rangle \big \langle {\a_{p}^{N_i}} \big | \phi_3 \phi_4 \big \rangle}{C_{p12}C^p{}_{34}} \Big ]_{\substack{h_i \ra \tilde{h}_i, \,h \ra \tilde{h}, \\ c \ra c -1}} \notag \\
  & = z^{\frac{(q_3 + q_4)^2}{k}} \F_p (c-1, \til{h}_i , \til{h}_p , z), \label{sugawarablock}
  }
where, for convenience, we suppressed the coordinate dependence of $\phi_i$ with respect to eq.~\eqref{vkmblock} and $\F_p (c-1, \til{h}_i , \til{h}_p , z)$ is the Virasoro conformal block in CFT, which in our conventions satisfies $\F_p (c-1, \til{h}_i , \til{h}_p , z) = z^{\til{h}_p} (1 + \dots)$.

Let us now consider the contribution of the $U(1)$ descendant sates to the Virasoro-Kac-Moody block. In what follows we will assume the level $k$ of the $U(1)$ algebra is positive so that all descendant states have positive norm. Recall from eq.~\eqref{virkmdescendants} that $\big | \til{\a}_{p}^{0,M_i}\big\rangle$ denotes a descendant state with a total number $M$ of $U(1)$ generators $P_{-m}$. We thus have
  \eq{
  \Ket{\til{\a}^{0,M_i}_{p}} = \K_{\{m_j, \s_j\}}^{-1/2} P_{-m_1}^{\s_1} \dots P_{-m_M}^{\s_M} \Ket{h_p,q_p}, \label{u1basis}
  }
where we assume $m_1 > m_2 > \dots > m_N$ in order to avoid overcounting states.

Thanks to the abelianness of the $U(1)$ algebra, the normalization $\K_{\{m_j, \s_j\}}$ is given by
  \eq{
  \K_{\{m_j, n_j\}} = \prod_{j=1}^{M} \( \frac{m_j \,k}{2}\)^{\s_j} \s_j! .\label{norm}
  }
This is a consequence of the anomaly and originates from terms of the form $P^{\s_1}_{m_1} P^{\s_1}_{-m_1}$ in the inner product for which $[P_n, P_m] = \frac{n k}{2} \,\d_{n+m}$, cf.~eq.~\eqref{algebra}. The factor of $\s_j!$ in eq.~\eqref{norm} accounts for the fact that we must commute each $P_{m_j}$ with $P_{-m_j}$ a total $\s_j!$ number of times when computing the norm of $P_{-m_j}^{\s_j}\Ket{h_p,q_p}$. Using eq.~\eqref{transformationprimaries} we find
  \eq{
  \big\langle{ \til{\a}^{0,M_i}_{p}} \big | { \phi_3 \phi_4}\big\rangle =  \K_{\{m_j, \s_j\}}^{-1/2} \prod_{j=1}^M  \big[q_3 (x^+_3)^{m_j} + q_4 (x^+_4)^{m_j}\big]^{\s_j} \BraKet{h_p,q_p}{\phi_3 \phi_4}{},
  }
and, likewise,
  \eq{
 \big\langle{\phi_1 \phi_2}  \big | {\til{\a}^{0,M_i}_{p}}\big\rangle = (-1)^M\K_{\{m_j, \s_j\}}^{-1/2} \prod_{j=1}^M \big [q_1 (x^+_1)^{-m_j} + q_2 (x^+_2)^{-m_j}\big]^{\s_j} \BraKet{\phi_1 \phi_2}{h_p,q_p}{},
  }
where we used the fact that $\sum_j \s_j = M$. Finally, from the definition of the Virasoro-Kac-Moody block in eqs.~\eqref{vkmblock} and~\eqref{productblock}, it follows that
  \eq{
  \V^{KM}_{p}(z) &= \sum_{M} \sum_{\{m_j, \s_j\}}    \lim_{x^+_1 \ra \infty} \frac{\big\langle{\phi_1(x_1^+) \phi_2(1)}  \big | {\til{\a}^{0,M_i}_{p}}\big\rangle \big\langle{ \til{\a}^{0,M_i}_{p}} \big | {\phi_3(z)\phi_4(0)}\big\rangle}{\BraKet{\phi_1(x^+_1)\phi_2(1)}{h_p,q_p}{}\BraKet{h_p,q_p}{\phi_3(z)\phi_4(0)}{}}, \notag \\
  &=  \sum_{M} \sum_{\{m_j, \s_j\}}  (-1)^M \prod_{j=1}^{M} \frac{1}{\s_j!} \( \frac{2q_2 q_3}{k} \frac{z^{m_j}}{m_j} \)^{\s_j}.
  }
The condition $m_1 > m_2 > \dots > m_N$ and the multinomial theorem guarantee that
  \eq{
   \V^{KM}_{p}(z) &= \sum_{M=0}^{\infty} \frac{1}{M!} \lb \log (1-z)^{2 q_2 q_3/k} \rb^M = (1-z)^{2q_2 q_3/k}. \label{KMblock}
  }
Thus,  the Virasoro-Kac-Moody block~\eqref{vkmblock} is given by
  \eq{
  \V_p(z) =  z^{\frac{(q_3 + q_4)^2}{k}} (1-z)^{\frac{2q_2 q_3}{k}} {\F} (c-1, \til{h}_i, \til{h}_{p},z).
  }
  %


\section{Mapping $\R_n$ to the complex plane in WCFT} \label{ap:map}

In this appendix we derive the map~\eqref{map} between the $n$-sheeted Riemann surface $\R_n$ and the complex plane $\C$. Our strategy is to consider the one-point functions of $T(\til{x}^+)$ and $P(\til{x}^+)$ on $\R_n$ obtained via the generalized Rindler method in~\cite{Castro:2015csg,Song:2016gtd}. We then find the warped conformal transformation that reproduces these one-point functions starting from the ones on the plane. This is the inverse of the approach used in~\cite{Calabrese:2004eu} to derive the conformal weight of the twist fields $\s_n$ in CFT.

Let $(\til{x}^+, \til{y}) = (\til{x}^+, \til{x}^- + \mu \log \til{x}^+)$ denote the spectrally-flowed coordinates on the $n$-sheeted Riemann surface ${\cal R}_n$ glued along the interval $\A$. The latter is defined as the line connecting the points $(0,0)$ and $(\ell, \bar{\ell})$ which we denote by
  \eq{
  \A:\quad (\til{x}^+, \til{x}^-) \in \big [ (0,0), \, (\ell,\bar{\ell}) \big].  \label{interval}
  }
The one-point functions of $T(\til{x}^+)$ and $P(\til{x}^+)$ on $\R_n$ are given by~\cite{Castro:2015csg,Song:2017czq},
  \eq{
  \Vev{T(\til{x}^+)}_{{\cal R}_n} & = \frac{\ell^2}{(\til{x}^+)^2 \big( \til{x}^+ - \ell \big)^2} \frac{h_n}{n}, 
  \qquad \Vev{P(\til{x}^+)}_{{\cal R}_n} = \frac{\ell}{\til{x}^+\big( \til{x}^+ - \ell \big)} \frac{i q_n}{n},  \label{TPRn}
  }
where $h_n$ and $q_n$ are the conformal weight and charge of the twist operator in the WCFT
  \eq{
  h_n = n \Big ( \frac{c}{24} + \frac{L_0^{vac}}{n^2} + \frac{i \a P_0^{vac}}{2\pi n} - \frac{\a^2 k}{16\pi^2} \Big), \qquad  q_n = \Big( P_0^{vac} + \frac{i n k \a}{4\pi} \Big). \label{twist}
  }

Following the discussion in Section~\ref{se:excited}, we identify the parameter $\a$ in eq.~\eqref{twist} with the spectral flow parameter $\mu$ of the WCFT via
  \eq{
  \mu = \frac{\a}{2\pi}.
  }
It follows that $L_0^{vac}$ and $P_0^{vac}$, the expectation values of $L_0$ and $P_0$ on the cylinder, satisfy
  \eq{
  L_0^{vac}  = -\frac{c}{24} -\frac{\mu^2  k}{4}, \qquad P_0^{vac} = -\frac{i \mu k }{2},
  }
which implies that the expectation values of the $T(x^+)$ and $P(x^+)$ currents on the spectrally-flowed plane with coordinates $(x^+, y) = (x^+, x^- + \mu \log x^+)$ are given by
  \eq{
    \Vev{T(x^+)} & = -\frac{k \mu^2}{4(x^+)^2}, \qquad \Vev{P(x^+)} = \frac{k \mu}{2 x^+}. \label{TPalpha}
  }

Using the transformation of the currents given in eqs.~\eqref{Tprime} and~\eqref{Pprime} we find that the following warped conformal transformation
  \eq{
  x^{+} = \bigg(\frac{\til{x}^+}{\til{x}^+ - \ell} \bigg)^{\frac{1}{n}}, \qquad y = \til{x}^{-} + \mu\log \bigg(\frac{\til{x}^+}{\til{x}^+ - \ell}\bigg), \label{mapder2}
  }
maps the expectation values of $T(x^+)$ and $P(x^+)$ on the spectrally-flowed plane~\eqref{TPalpha} to the corresponding expressions on $\R_n$~\eqref{TPRn}. Alternatively, the transformation
  \eq{
  x^{+} = \bigg(\frac{\til{x}^+}{\til{x}^+ - \ell} \bigg)^{\frac{1}{n}}, \qquad x^{-} = \til{x}^{-} + \mu \bigg(\frac{n-1}{n}\bigg) \log \bigg(\frac{\til{x}^+}{\til{x}^+ - \ell}\bigg) , \label{mapder1}
  }
maps the expectation values  on the reference plane, eq.~\eqref{TPalpha} with $\mu = 0$, to the expectation values on $\R_n$ given in~\eqref{TPRn}. 

Note that the warped conformal transformations~\eqref{mapder2} and~\eqref{mapder1} are independent of the length of the interval $\A$ along the $\til{x}^-$ coordinate. This is a feature of the Rindler map used to derive the entanglement entropy in WCFT and is also reflected in eq.~\eqref{TPRn}.




\ifprstyle
	\bibliographystyle{apsrev4-1}
\else
	\bibliographystyle{jhep}
\fi

\bibliography{chaoswcft}



\end{document}
